\documentclass[12pt,twoside]{article}
\usepackage{epsfig}
\newcommand{\be}{\begin{equation}}
\newcommand{\ee}{\end{equation}}

\begin{document}
\title{\bf Tunneling in very small GaAs MESFET }
\author{W. Poirier, D. Mailly * and  M. Sanquer **}
\maketitle
\begin{center}
\small CEA-DSM-DRECAM-SPEC,
C.E. Saclay,  91191 Gif sur Yvette Cedex,  * CNRS-LMM, 196 Ave. H. Ravera, 92220 Bagneux, France, and ** CEA-DSM-DRFMC-SPSMS, CEA-Grenoble 17 rue des martyrs 38054 Grenoble, France.
\end{center}
\begin{abstract}
We study the transport through gated GaAs:Si wires of 0.5 micrometer length in the insulating regime and observe transport via elastic tunneling at very low temperature. We describe the mean positive magnetoconductance and the mesoscopic fluctuations of the conductance ( versus energy or magnetic field ) purely within one electron model without introducing Coulomb blockade considerations.\\
$ $ 
\end{abstract}
\par During the eighties, mesoscopic Si MOSFET's have been extensively studied both in the insulating and diffusive regimes. The huge reproducible conductance pattern observed when one varies the carriers energy in the insulating regime, has been qualitatively understood both in short and long samples \cite{fowler}. In long samples, conductance fluctuations are the result of an incoherent mesoscopic effect appearing in the variable range hopping regime \cite{lee}\cite{raikh}: the dominating resistor in the sequence of hops fluctuates with the energy. Recently this effect has been theoretically as well as quantitatively studied by various authors \cite{ladieu}\cite{hughes}\cite{orlov}. On the other hand, in the shortest Si MOSFET samples (~ $0.5 \mu m$, shorter than the averaged Mott hopping length at low temperature~), resonant tunneling has been  observed \cite{fowler}. Some debate has occurred to know the role of inelastic processes to explain the amplitude and the temperature dependence of the resonant conductance peaks observed in gate voltage \cite{stone}\cite{azbel}. 
\par At the beginning of the nineties, attention has been paid on the role of Coulomb blockade in disordered insulating wires \cite{chandrasekhar}\cite{scott-thomas}\cite{meir}\cite{johnson}: conductance resonances periodic in gate voltage have been observed, near the metal-insulator transition in simple disordered wires or in disordered cavities. After this time,
Coulomb blockade effects have been often invoked in small gated disordered semiconducting wires near the pinchoff voltage to explain the observed conductance fluctuations \cite{degraaf}\cite{leobandung}\cite{hiramoto}\cite{nakazato}\cite{staring}\cite{imam}, even if in general no periodicity in gate voltage is observed.  The aperiocicity can be obtained for Coulomb blockade implying few "islands" of various capacities in series in the sample.
\par The role of electron interactions in disordered insulators is indeed a very old unresolved question. The Coulomb interaction can manifest itself in different ways: First it can produce a so-called "Coulomb gap" \cite{shklovskii}, in the case where the charging energy ${e^{2}\over \epsilon \xi}$ is lower than the mean level spacing in a localization domain: $ \Delta \simeq {1 \over n(E_{F}) \xi^{d}}$. This means that the charge neutrality on size of the localization domains is not perfect. Secondly, it can induce the Coulomb Blockade regime when the charging energy is the dominant one (that supposes inhomogeneity). Finally, it has been recently claimed that (~repulsive or attractive~) interaction can  induce delocalization of electrons \cite{shepelianski}. For these reasons, the study of mesoscopic insulators near the  Metal-Insulator transition, and the separation between independent and interacting electrons picture is of great interest.

\par It is consequently an important challenge  to know if the insulating state of small disordered semiconducting wires is due to an {\it interference effect}, as described by independent electron models of an Anderson insulator, or to a {\it charging effect} (more generally an interaction effect), as described by the Coulomb blockade if inhomogeneities are postulated.
\par Inhomogeneity of the disorder configuration or sample shape is an essential ingredient for Coulomb blockade interpretation: a  (or few) metallic "island" sharing many electrons is supposed to be separated from the rest of the wire by strongly insulating regions acting as capacitor dielectrics \cite{staring}. Such picture can explain  periodic conductance resonances in gate voltage corresponding to the addition of one electron in the island. The activation energy for the conductance out of resonance is determined by the charging energy $e^{2}\over C$ where $C$ is the total capacitance of the island ( to the gate and to the rest of the wire) which  exceeds the bare mean level spacing $\Delta E$. This gives a way to explain large activation energies as compared to the bare level spacing. 
\par In the Coulomb blockade model, the magnetic field does not
 decorrelate the pattern of conductance peaks seen in gate voltage in the insulating regime, in contrast with the case of Universal Conductance Fluctuations (~UCF~) in the diffusive regime. This is often considered as a sign for Coulomb blockade, although a mesoscopic wire of a disordered Anderson insulator (~independent electron model~) exhibits also
a basic non ergocicity in magnetic field.
\par Moreover the so-called "Coulomb staircase" is not observed in disordered semiconducting wires, but only large I-V non-linearities, also attempted in disordered insulators without interaction.
\par In this paper we will  systematically use independent electron models to interpret our results on mesoscopic GaAs:Si wires. At this stage we do not have to incorporate Coulomb blockade effect to explain the main observed features. 
\par In contrast to Coulomb blockade analysis, our interpretation is {\it quantitative} for the amplitude of the conductance fluctuation as well as for the mean magnetoconductance effects. We also understand the whole gate voltage range, from the deeply localized regime to the diffusive regime, and in particular the occurrence of the metal-insulator transition, within the Anderson scheme.  So one aim of this work is to show that in disordered insulating wires, when periodic oscillations in gate voltage are not seen, independent electron models have to be considered first, in contrast to the case of intentionnally build cavities.
\par Like Si MOSFET, gated GaAs:Si  wires permit to tune the Metal-Insulator transition with gate voltage.
 We present conductance measurements in sub-micronic GaAs MESFET at very low temperature near the metal-insulator transition and in the insulating regime. The sample are very small and studied down to very low temperature, that permits us to enter in the elastic transport regime, in contrast to previous studies \cite{ladieu}\cite{hughes}. As compared to  short Si MOSFET's studied by Fowler et al. \cite{fowler}, our quasi one dimensionnal samples are smaller than $1 \mu m$ in all dimensions. We have been able to see the appearance of very high conductance resonances superimposed to a background of low monotonic mean conductance variation, first described theoretically by Azbel and coworkers in simulation of 1D disordered wires near zero temperature \cite{azbel}. This is in contrast to the Gaussian distribution for the logarithm of the conductances observed in longer quasi1D samples, which results from the truncation of quantum fluctuations by geometrical incoherent mesoscopic effect \cite{lee}\cite{raikh}\cite{ladieu}\cite{hughes}.
\par Our short samples departs from long samples in numerous ways, which will be considered in the following:
\begin{itemize}
\item There exists very pronounced conductance resonances in energy. The histogram of conductance exhibits a long tail to high conductance values deep in the insulating regime. 
\item The temperature dependence is ${\delta G \over \delta T} \le 0$ for large  resonances.
\item  The amplitude of the conductance fluctuations is close to its quantum zero temperature estimation.
\item The mean spacing in energy between successive maximum of $\ln (G)$ shows some spacing rigidity. 
\end{itemize}
\par As said before the magnetoconductance is  important to test the role of quantum interferences in  insulating wires. In our sample the role of the magnetic field is fully understood within independent electrons models. For a disordered  Anderson  quasi 1d coherent insulator, the conductance is given by:
\be G \simeq {e^{2}\over h} \exp (- 2 L / \xi) \ee with 
\be \xi = ( \beta {\cal N} + 2 - \beta ) \ell \ee
 (~$\beta  $ is a basic symmetry dependent parameter~) \cite{pichard}. Equation (1), obtained from random matrices theory arguments, shows that the localization length $\xi$ is symmetry-dependent (~through $\beta$~) and depends on ${\cal N}$.  $\beta = 1 , 2 , 4 $ respectively for the orthogonal ( time reversal + spin reversal symmetry ), unitary ( no time reversal symmetry) and symplectic (time reversal symmetry only) cases. 
The $\beta $ dependence has been observed experimentally by various authors for large samples for which the number of channels ${\cal N}$ is large \cite{pichard}\cite{gershenson}: In  quasi-1D systems without spin-orbit scattering the localization length is doubled by application of a flux quantum through the localization domain. This implies a very large positive magnetoconductance whose importance has only be recognized recently. The ${\cal N}$ dependence has not been tested up to now. Our samples offer this opportunity (see section magnetoconductance). 
 Equation (1) shows that for ${\cal N} = 1$, the localization length is reduced to twice the elastic mean free path $\ell$,
and there is no $\beta$ dependence of $\xi$. From the deeply localized regime, by increasing the gate voltage, one increases  ${\cal N}$, which is about 8 in our sample at the metal-insulator transition (see next section). For ${\cal N} \ge 1$, $\xi$ is larger than $\ell$ and should increase strongly by application of a magnetic field ($\beta = 1$ to $\beta = 2$  ).

\section{samples and experiment} 
\begin{figure}[ptb]
\begin{center}
\epsfig{file=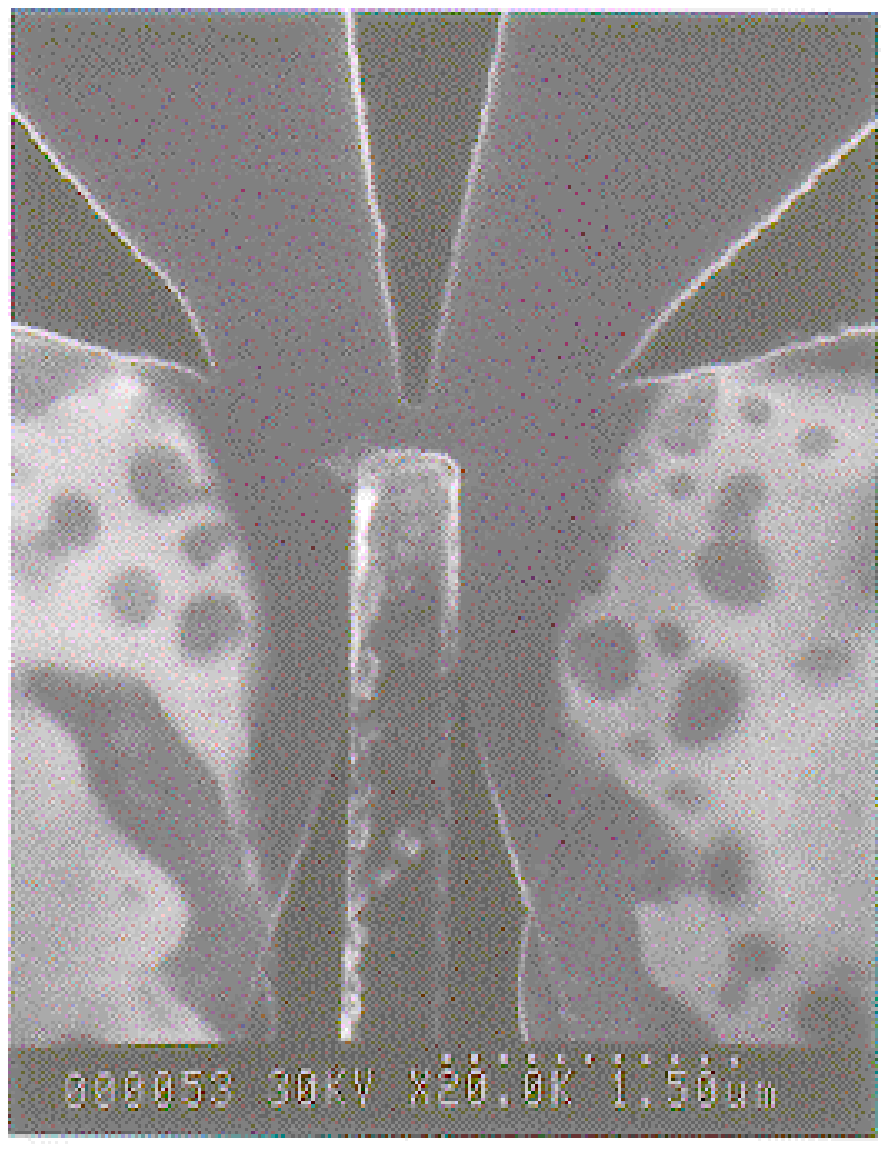,width=14cm}
\caption{  SEM picture of the GaAs:Si submicronic MESFET. The 0.5 micrometer thick Aluminium Schottky  gate is  visible on the bottom. The GaAs (doped at $10^{23} Si \ \ m^{-3}$) 300nm thick layer is  etched  to form 4  large contact pads to the active region under the gate. AuGeNi ohmic contacts are visible on the right and the left. The volume of the active region is estimated to be $0.2 \times 0.2 \times 0.5\mu m^{3}$ (~taking into account depletion layers for $V_{G}=0V$~).}
\label{fig:echantillon}
\end{center}
\end{figure}

 Figure \ref{fig:echantillon} is a SEM picture of our smallest MESFET (~Metal-Semiconductor Field Effect Transistor~). The sample has a form of a constriction defined by etching and  its central part is covered by a Schottky Aluminium gate. The section of the etched region grows rapidly to large contacts, such that the resistance is dominated
by the region under the gate. Longer samples (~not shown~) have been also studied. The concentration of Si dopands (~$10^{23} Si \ \ m^{-3}$ on a thickness of 300nm~) is choosen to be close to the metal-insulator transition but on the metallic side 
with zero gate voltage. The section (~defined by etching~) is larger than the depletion region at the surface.
The sample is placed inside a dilution refrigerator equiped with coaxial cables. Proper filtering is used at the top of the refrigerator. The conductance  at few resonances increases between
 $T=100 mK$ and   $T=30 mK$ (~see figure \ref{fig:peakdet}~), that proves the good thermalization of  electrons. DC and AC (~33 Hz~) measurements with small excitation levels are used depending on the resistance range. 
\par
\begin{figure}[ptb]
\begin{center}
 \epsfig{file=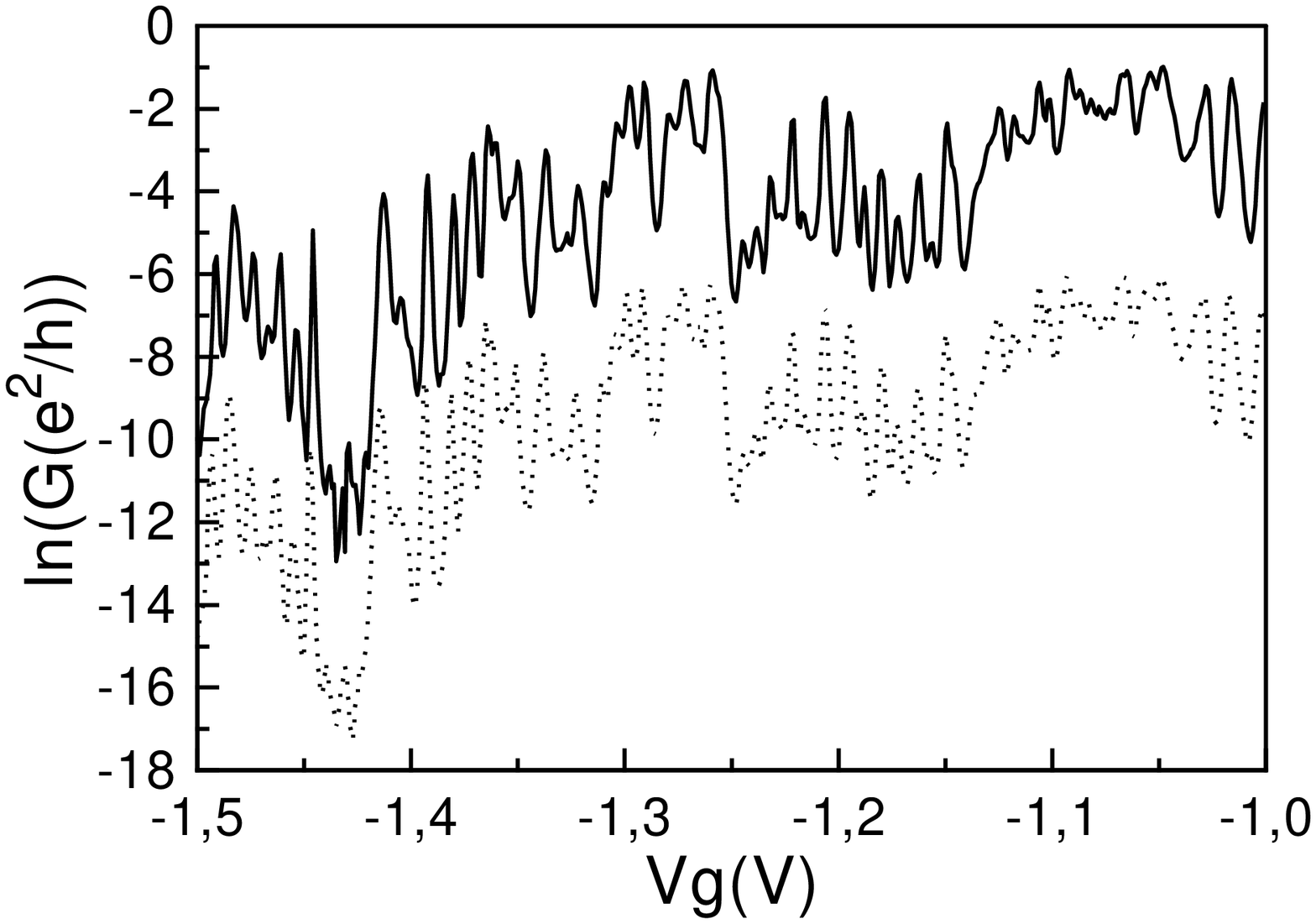, width=10cm}
\caption{$\ln$G(Vgate) at T=100mK in the $0.5 \mu m$ long sample for two successive experiments without thermal cycling, showing the excellent reproducibility of the conductance pattern.}
\label{fig:reproducibility}
\end{center}
\end{figure}

\begin{figure}[ptb]
\begin{center}
 \epsfig{file=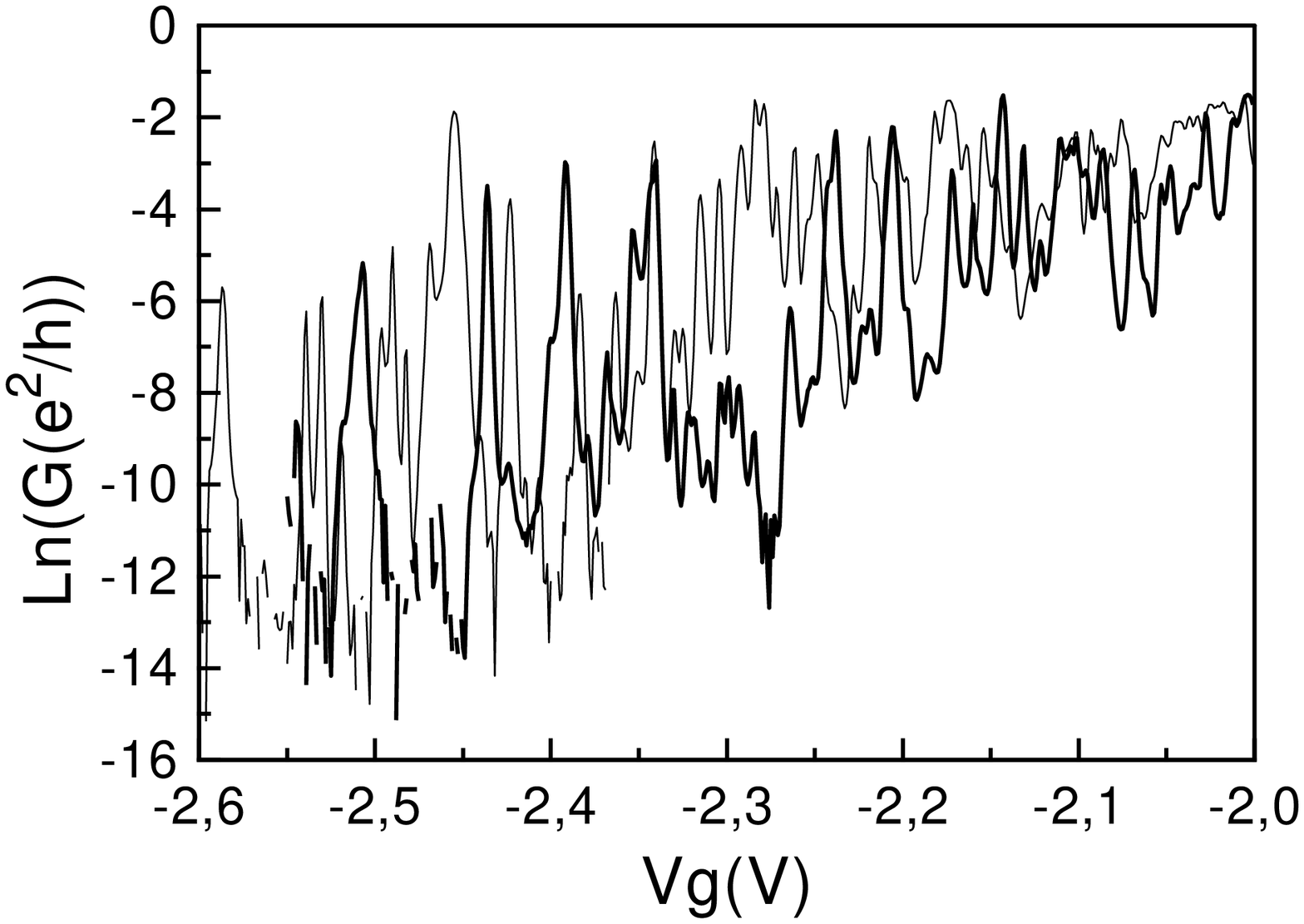,width=10cm}
\caption{$\ln$G(Vgate) at T=100mK in the $0.5 \mu m$ long sample, for two successive experiments with a thermal cycling to room temperature, showing a complete decorrelation. The pattern of conductance is sensitive to the details of the disorder configuration.}
\label{fig:annealing}
\end{center}
\end{figure}

  Figure \ref{fig:reproducibility} shows the excellent reproducibility of the conductance pattern versus gate voltage between successive experiments performed at very low temperature. Such reproducibility is obtained as long as the sample is kept at low temperature, and after a stabilization of the gate by few gate voltage cyclings. On the contrary, figure \ref{fig:annealing} shows that annealing at room temperature completely decorrelates the pattern of conductance. This indicates that the disorder potential is not related to structural inhomogenities, like large fluctuations in the dopand concentration or
in thickness of the wire.
\par 

\par
\begin{figure}[ptb]
\begin{center}
 \epsfig{file=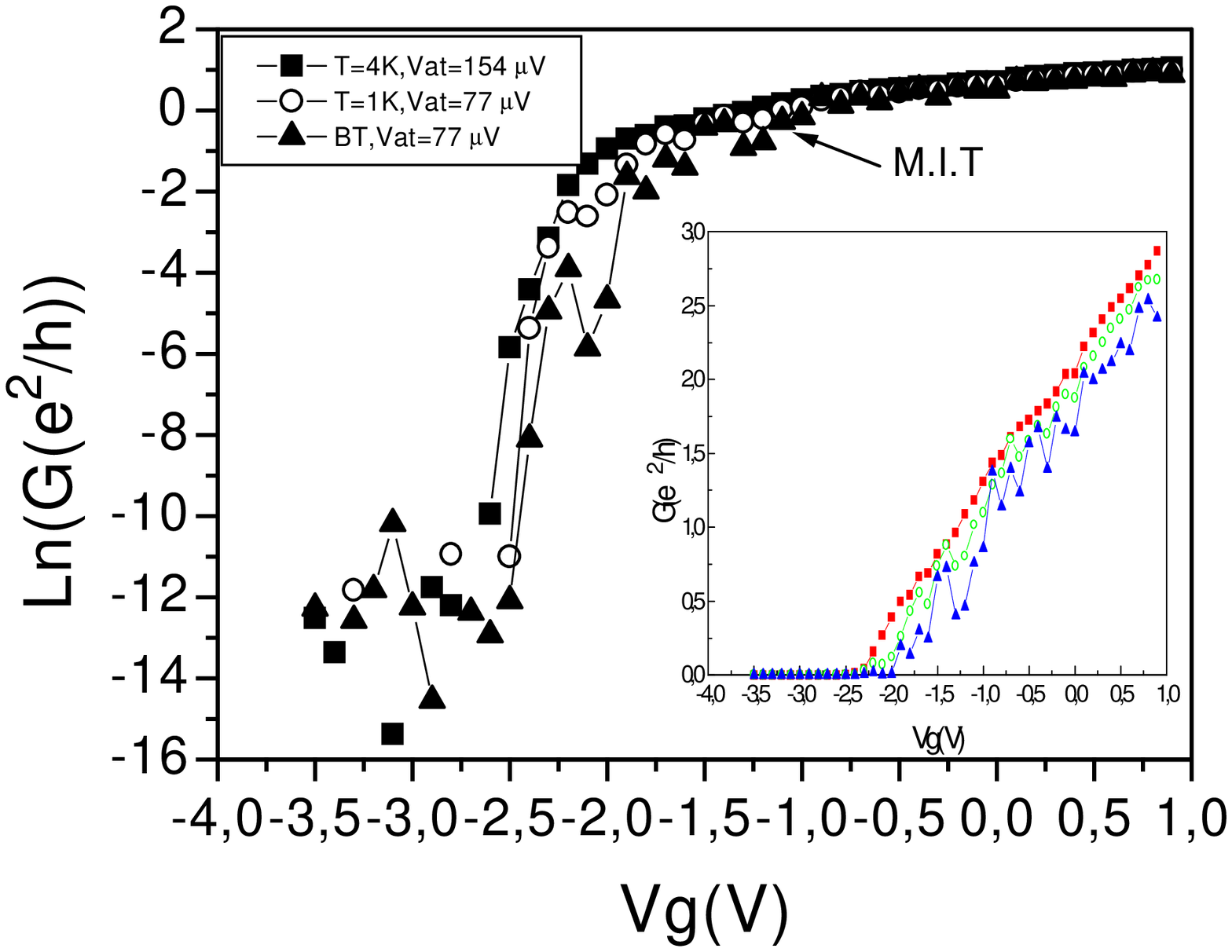,width=10cm}
\caption{$\ln$G(Vgate) at three temperatures in a large gate voltage range (details of the conductance pattern are not seen vor this gate voltage sampling). Inset: the same curve in a linear scale. Note the linearity at voltages above the transition.}
\label{fig:transition}
\end{center}
\end{figure}

 As one decreases the gate voltage, the wire is progressively depleted. In the metallic regime, the conductance varies linearly with $V_{G}$ as shown on figure \ref{fig:transition}. In this regime the cross section of the wire is reduced as $V_{G}$ decreases, with an approximately constant density of levels $n_{e} \simeq 10^{23} m^{-3}$ in the wire (~this corresponds to a perfect screening of the gate induced potential change in the middle of the wire~).
The Boltzman conductance of the wire is given by
\be G = {n_{e} e^{2}\tau \over m^{*}}{W^{2}\over L } = {N_{e} e^{2}\ell \over \hbar k_{F}}{1\over L ^{2}} =  {\pi \over 3}{e^{2} \over h} {\cal N} { \ell \over L}
\ee
 where $W^{2}$ is the cross section of the wire, $L$ is the length of the wire, $k_{F} = (3 \pi^{2} n_{e})^{1/3}$ is the Fermi wave vector,  $N_{e} = n_{e} L W^{2} $ is the number of electrons in the wire, and   ${\cal N} = ( {k_{F}W \over \pi})^{2}$ is
the number of occupied 1D subbands (~or channels~).
The measured conductance is: \be G \simeq 8.3 \ \ 10^{-5} ( 1 + { V_{G}\over 2.5})\ee
Taking the elastic mean free path $\ell $ to be the distance between Silicon dopants
(~20 nm~), and supposing that there is no more
electrons for $V_{G} \simeq -2.5 V$, we obtain the number of electrons in
our wire as a function of $V_{G}$ (~in volts~):
\be N_{e} \simeq 615 ( 1 + { V_{G} \over 2.5})
\ee 
This estimation  corresponds also to a gate capacitance of $4 \ \ 10^{-17} F$, in perfect accordance with $C = {\epsilon_{0}\epsilon_{r}S \over {\cal L}}$ (~$S =  W \times L = 70 nm \times 0.5 \mu m$ is the typical surface of the sample near the transition (~see later~), ${\cal L} = 120 nm$ is the depletion length and $\epsilon_{r} = 12$ in GaAs~). The number of electrons is about  400 in the wire at $V_{G} \simeq -1 V$. 

\par
\begin{figure}[ptb]
\begin{center}
 \epsfig{file=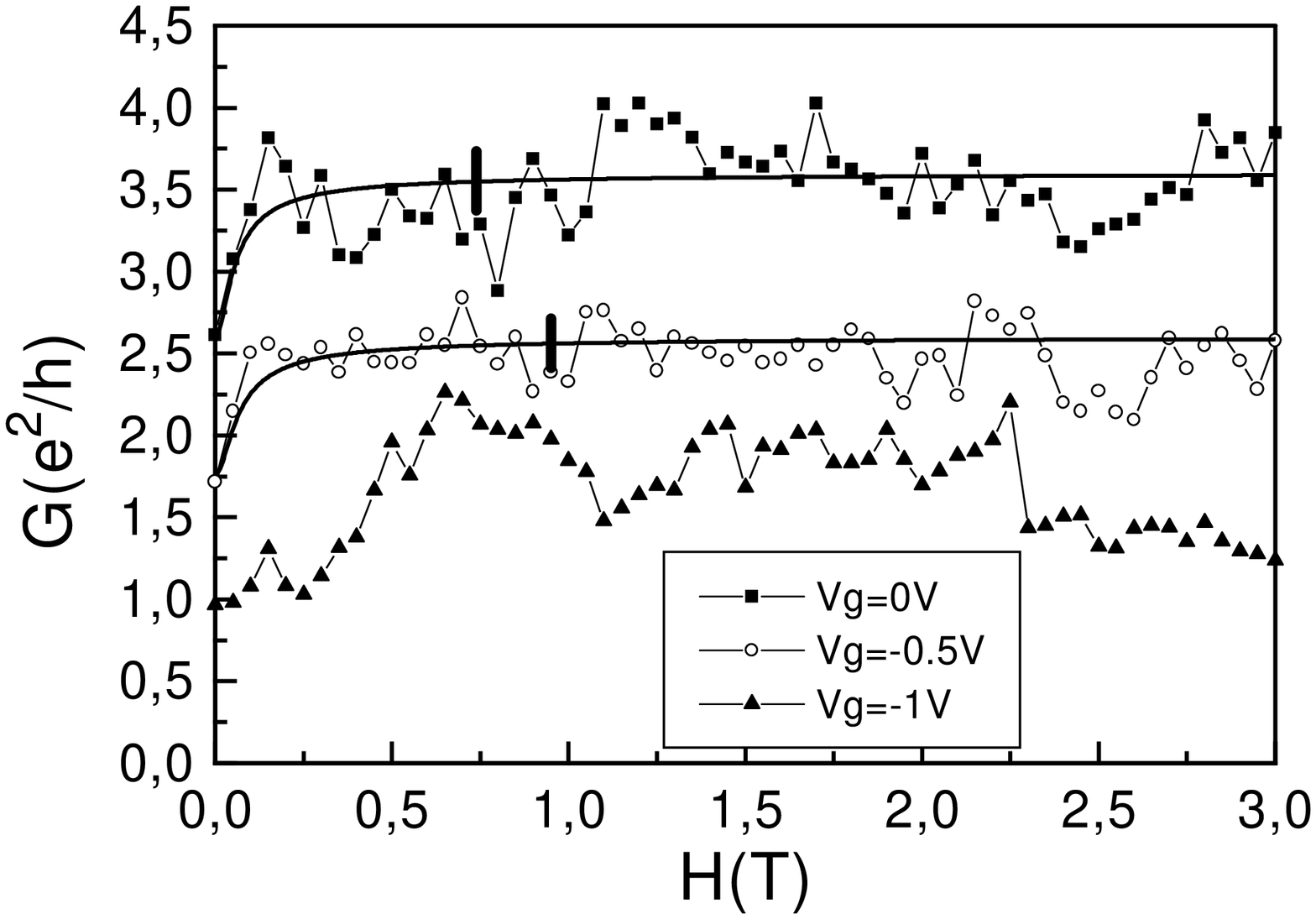,width=10cm}
\caption{The magnetoconductance at T=100mK for three gate voltages. In the diffusive regime above the metal-insulator transition the positive weak field magnetoconductance is well fitted by the weak localization  term ( the phase breaking length is about $0.25 \mu m$ for $V_{G}= 0 V$ and $0.21 \mu m$ for $V_{G}= -0.5 V$ (thick lines)). Note the qualitative change for $V_{G}= -1.0 V$ which is on the insulating side of the transition ( $G \simeq {e^{2}\over h}$). The vertical bars are the estimated UCF amplitude in the diffusive regime.}
\label{fig:mcmit}
\end{center}
\end{figure}

 The Metal-Insulator Transition (MIT), which is identified by a change in the magnetoconductance
curves (~see figure \ref{fig:mcmit}~), takes place at $V_{G} \simeq -1 V$
(~in our short mesoscopic samples it is impossible to detect the transition by a fine analyse for the temperature dependence of the conductance~). For $V_{G} = -0.5V$ and for $V_{G} = 0V$, it is possible to fit the magnetoconductance by weak-localization formula, which gives
$ W \simeq 0.12 \mu m$ for the width of the wire, and $ L_{\phi} \simeq 0.2 \mu m \simeq L/2$ for the phase breaking length.
\par At  $V_{G} \simeq -1 V$, when the change in the low field magnetoconductance happens, the measured conductance is approximately ${ e^ 2 \over h} $.    
A perfectly coherent sample at the transition (~$\xi$, the localization length $\simeq L \le L_{\phi}$~) should have a conductance $G = { 2e^ 2 \over h}$. This factor of two of departure  is compatible with our estimation of the phase breaking length at the MIT, which is less than the sample length: for $V_{G} \simeq -1 V$, the conductance is $G = {L_{\phi}\over L}{ 2e^ 2 \over h} 
\simeq { e^ 2 \over h} $.

\par  The  cross section
near the transition is estimated to be:
$ W^{2} = 70 nm \times 70 nm$, that corresponds to about ${\cal N} \simeq 10$.
\par {\it It is important to realize that  our disordered wires exhibit their MIT for a large number of channels, in contrast to the case of ballistic point contacts. Consequently near the transition in the insulating regime, the localization length is much larger than the mean free path (~Anderson insulator~).}

\par Of course near the transition, the hypothesis of a perfect screening and a "side depletion" is less valid. In the insulating state, the depletion due to more negative gate voltages is a bulk effect and $k_{F}$ decreases. Near the transition for instance, the Ioffe-Regel criteria $k_{F} \ell \simeq \pi$
should be roughly obeyed. It is  still possible to estimate the number of channels ${\cal N}$ (and consequently the localization length $\xi$) as function of the gate voltage in this regime.
Keeping the same gate voltage dependence for $N_{e}$, and noting that $N_{e}$ varies like  $k_{F}^{3}$ and  ${\cal N}$ like $k_{F}^{2}$, we found:
\be {\cal N}(V_{g}) \simeq  11( 1 + {V_{g} \over 2.5})^{2/3} \ee 
for $V_{g} \le -1 $.
 For instance, there is only one canal left for $V_{g} = -2.4 V$. For this value, the localization length is reduced to twice the elastic mean free path $\ell$,
and there is no more symmetry dependence of $\xi$ (~equation (1) with ${\cal N}=1$~). For
 $V_{g} = -1.0 V$ on the contrary, ${\cal N}$ = 8 to 10, and $\xi$ is much larger than $\ell$:
$\xi \simeq (N + 1)\ell \simeq 0.18 \mu m$.

\par The relation between $V_{G}$ and the energy of electrons is given by:
${\Delta E  \over \Delta V_{G}} \simeq 140 mK ( mV)^{-1}$ (~ near the transition, supposing a parabolic band~). 
 In the insulating regime, the typical spacing between conductance resonances is
$\Delta V_{G} \simeq 7 mV$ for the short sample, i.e. $\Delta E \simeq 1 K$, a realistic value for the separation between energy levels in the whole sample. In fact, supposing a constant sample volume and a constant density of states in the impurity band up to the Fermi level at the MIT, which occurs for a Fermi energy of about 200K (Ioffe-Regel criteria), we estimate $\Delta E  \simeq { 200K \over L W^{2}n_{e} } \simeq 0.8 K$~). We observe that this spacing is roughly constant in the whole gate voltage range between $V_{G}= -2.5V$ and  $V_{G}= -1V$  
\par We can also estimate the Mott hopping length on the insulating side of the transition (~although we cannot rely on the temperature dependence in our mesoscopic sample~): $r_{Mott} = \xi ({T_{0}\over T})^{1/2}$ with $T_{0} = { 1 \over k_{B}  \xi W^{2}n_{e}} \simeq 2.3K$.
\par This gives $<r_{Mott}> \simeq 0.27 \mu m $ at the transition for $T=1K$ (with our estimation of $\xi$). As shown by Ladieu et al. \cite{ladieu}, for this quasi 1D geometry, one has better to consider $<r_{Max}> = <r_{Mott}> \sqrt{2 \ln (a {L \over<r_{Mott}> })}
\simeq 0.51 \mu m$ ($a \simeq 2$), comparable to the sample length $ L \simeq 0.5 \mu m$.
\par {\it This indicates that near the transition the estimations for $L_{\phi}$,
$\xi$ and $r_{Mott}$ converge to the sample length in our short sample.} Due to the small dependence of $r_{Mott}$ on energy, even in the deep insulating regime (~i.e. when $\xi$ decreases down to 55 nm - its estimation at $V_{G}=-2.4V$ -~), the Mott hopping length is comparable to the short sample length. 
{\it This means that the transport should be essentially elastic.}

\section{Temperature dependence in the resonant tunneling regime}

\par
\begin{figure}[ptb]
\begin{center}
 \epsfig{file=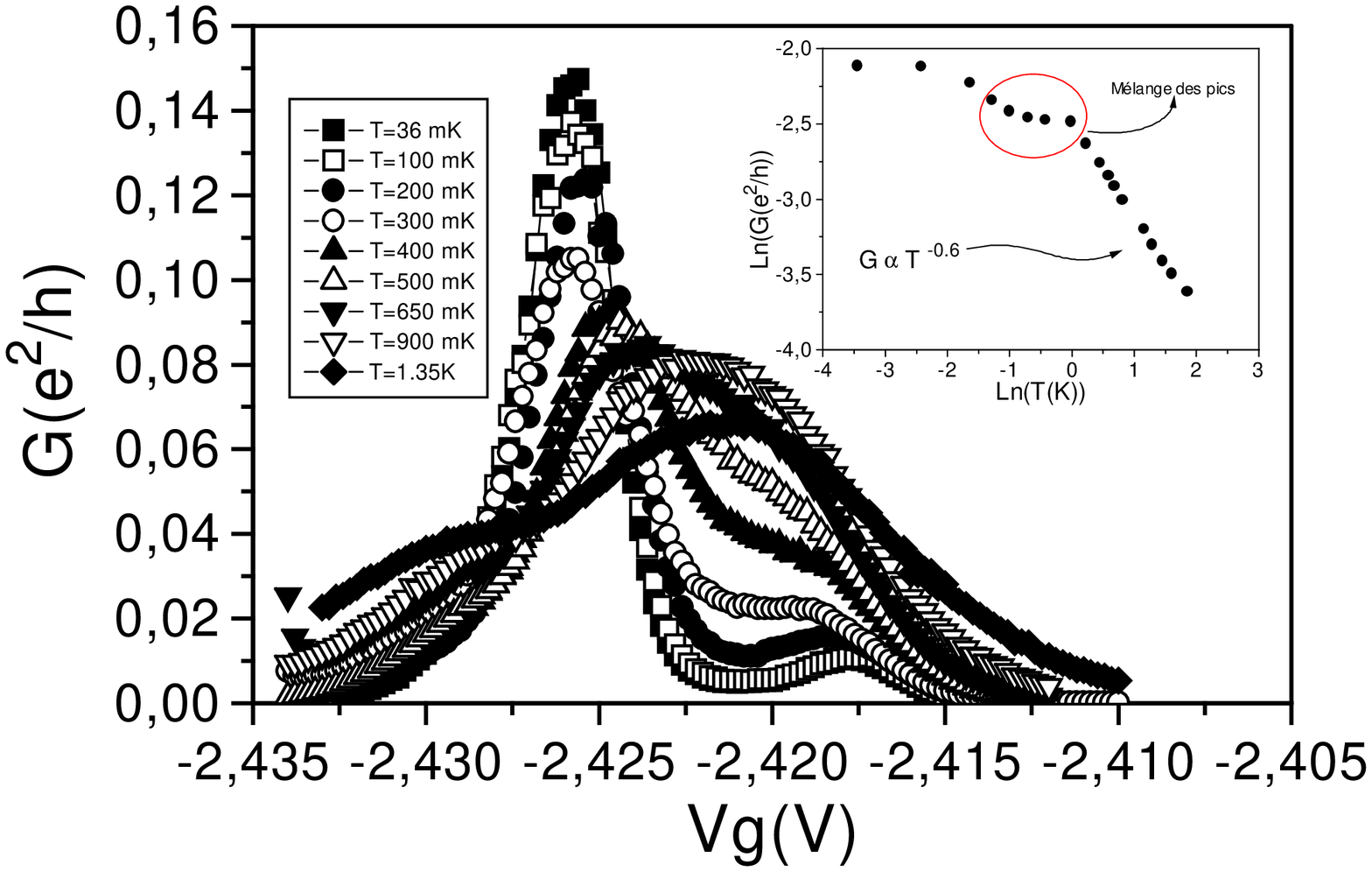,width=11cm}
\caption{ Detail of G(Vgate) in the $0.5 \mu m$ long sample, showing two resonant peaks at various temperatures in the deep localized regime.}
\label{fig:peakdet}
\end{center}
\end{figure}

\begin{figure}[ptb]
\begin{center}
 \epsfig{file=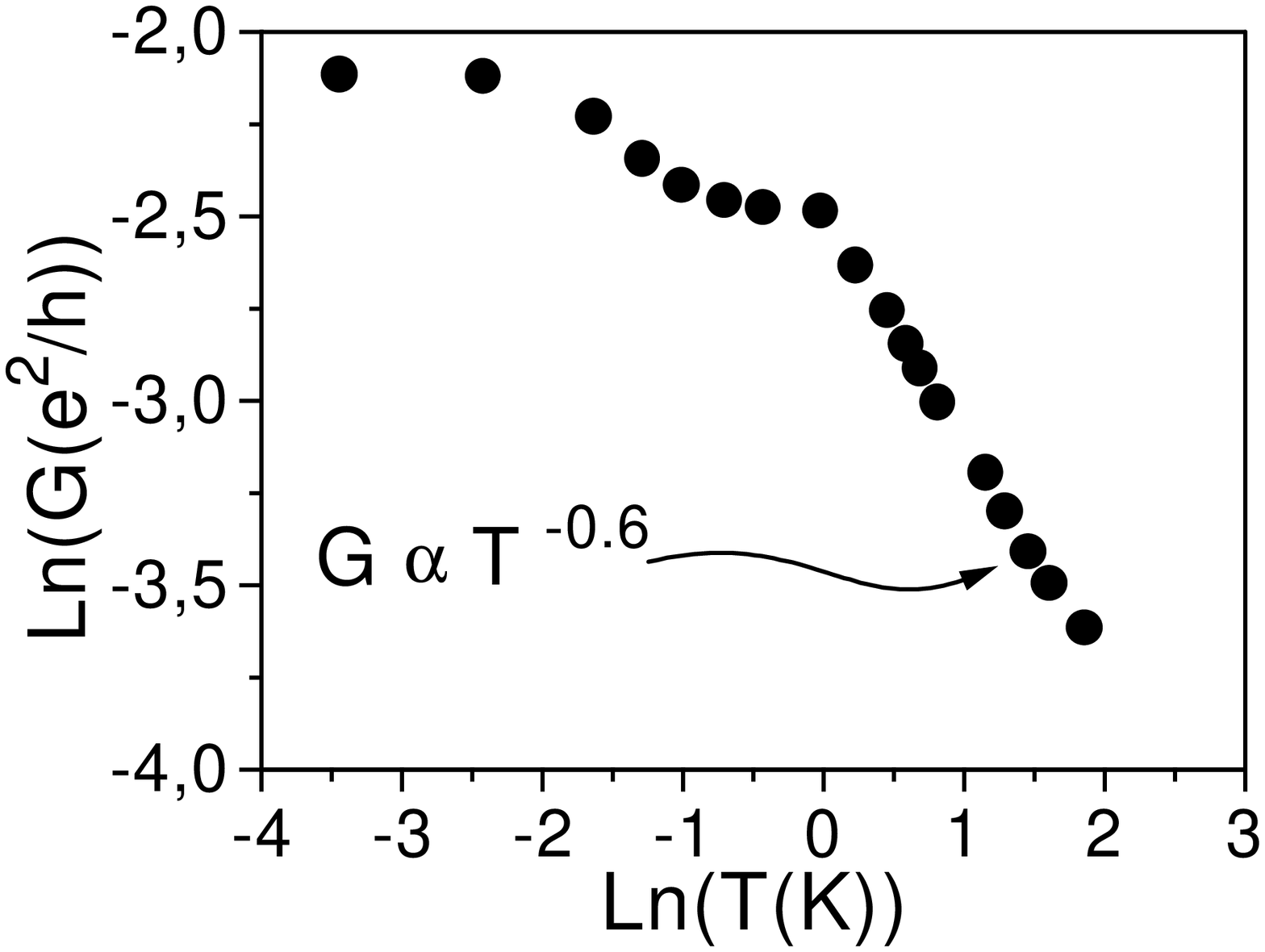,width=11cm}
\caption{G(T) at the resonance (~see figure 6~). Note that ${\delta G \over \delta T} \le 0$ up to $T=7K$.}
\label{fig:gdet}
\end{center}
\end{figure}

 Indeed, contrary to the widely reported case of long samples, we observe in our short samples  (length $\le 0.5 \mu m$) that  for few reproducible peaks of conductance in gate voltage, the conductance {\it increases} at very low temperature (typically one order of magnitude between T=35mK and T=4.2K) (~see figures  \ref{fig:peakdet} and \ref{fig:gdet}~),. This characterizes {\it resonant } transport through the disordered insulator at very low temperature. The conductance at the resonance is sometimes larger than 0.1 quantum of conductance, even far in the insulating regime, where the log-averaged conductance in gate voltage gives typically $ G \simeq 10^{-4}$ in quantum units.
\par The distinguishing feature for  short  wires (~as compared to the long ones~) is the existence of these very large conductance peaks.
 \par In wires whose length is comparable to the phase breaking length, the transport is by tunneling. Direct tunneling between reservoirs is negligible except for very short samples. The dominant conductance mechanism is the resonant tunneling via a localized state (~a single impurity site in the deep insulating regime or an extended localized state close to the MIT~) near the center of the wire, or via a serie of equidistant localized states of the same energy \cite{azbel}.

The temperature dependence of the conductance for resonant tunneling through a single localized state has been considered by various authors \cite{azbel}\cite{stone}.
Figure  \ref{fig:peakdet} shows the evolution in temperature for two nearest conductance peaks (~separated by approximately 1 kelvin~) in the deeply localized regime. For temperatures larger than 0.9K, the two peaks merge into a single broad resonance whose width is comparable to the temperature (~${\Delta E  \over \Delta V_{G}} \simeq 140 mK ( mV)^{-1}$~). This is the high temperature classical regime. Below $T \simeq 0.9K$, the two peaks become  resolved in gate voltage. The width of the peaks saturates to $\delta E \simeq 0.45K$ below  $T \simeq 0.3K$. We suppose that  $\delta E  $ corresponds to the inelastic rate $\Gamma_{i}$
(~at this temperature~). This value is close to  the inelastic rate estimation in the diffusive regime near the transition (~$\tau_{\phi} \simeq 17 ps$~). The temperature regime $ k_{B}T \ge \Gamma_{i}$ is called the thermally broadened regime \cite{meirav}, and the conductance at the resonance should follow a $1/T$ dependence. On figure \ref{fig:peakdet} we observe a smaller exponent: $ G \simeq T^{-0.6}$. For temperatures comparable or smaller than $  \Gamma_{i}$, the value of the conductance at the resonance is given by:
\be G( E=E_{res}) =  { 2 e^{2}\over h}  { \Gamma_{e}\over \Gamma_{i}} 
\ee
where $\Gamma_{e}$ is the elastic broadening. The temperature dependence should saturate (~except if  $\Gamma_{i}$ itself depends on the temperature~) as roughly observed in  figure \ref{fig:gdet}.
 We observe for the largest peak (~see figure  \ref{fig:peakdet}~) that $G( E=E_{res}) \simeq 0.15{  e^{2}\over h} $, i.e.  $\Gamma_{e} \simeq 0.035 mK$.
Such a large value of $\Gamma_{e}$ means a very large fluctuation of the localization length because
\cite{stone}:
\be \Gamma_{e} = {\Delta E \over 2\pi} \exp (-{L \over \xi})
\ee
that gives $\xi \simeq 0.3 \mu m$. Such fluctuations are seen in zero temperature simulations of independent electron transport in disordered insulators (~see later~).

\par
\begin{figure}[ptb]
\begin{center}
 \epsfig{file=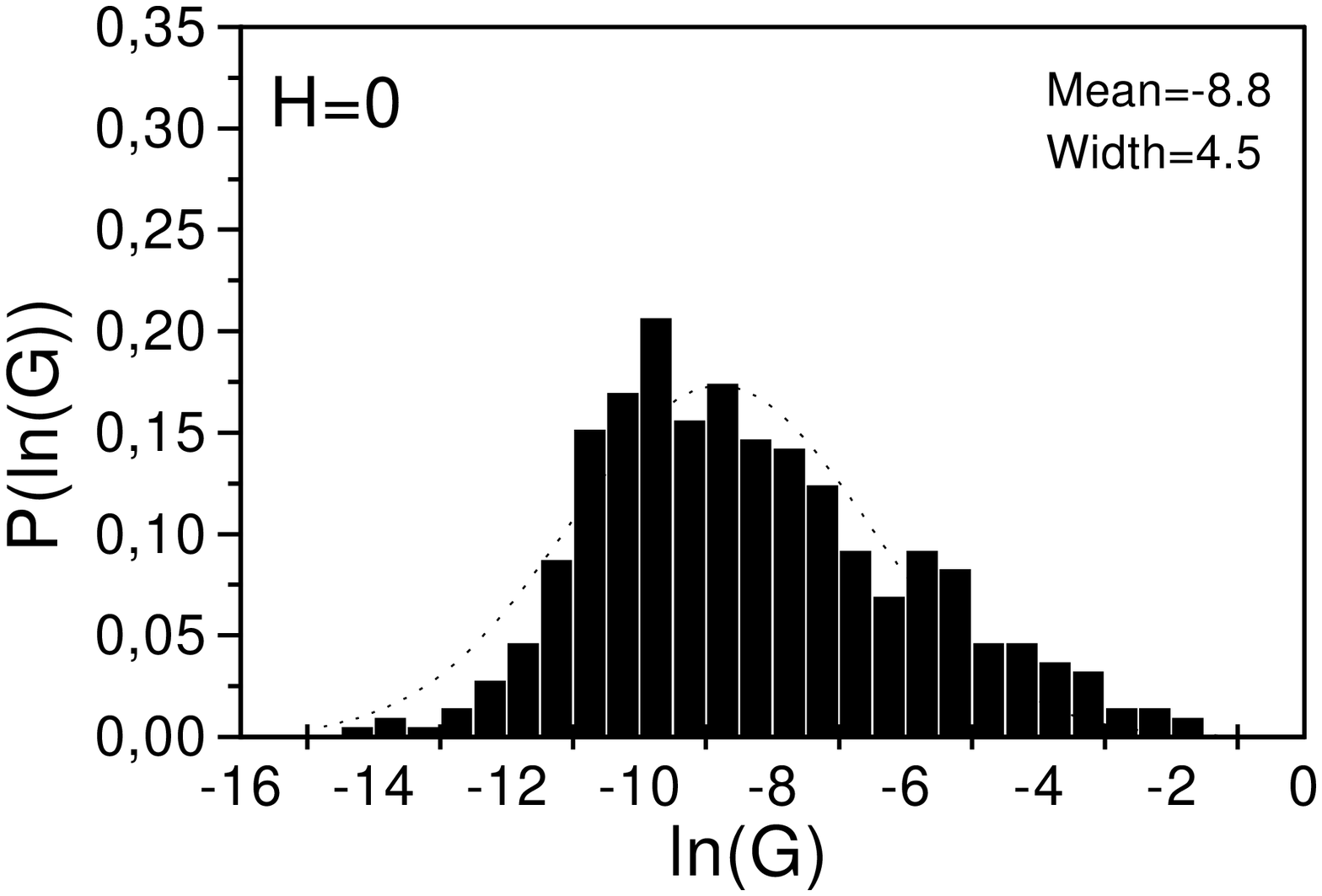,width=8cm}
 \epsfig{file=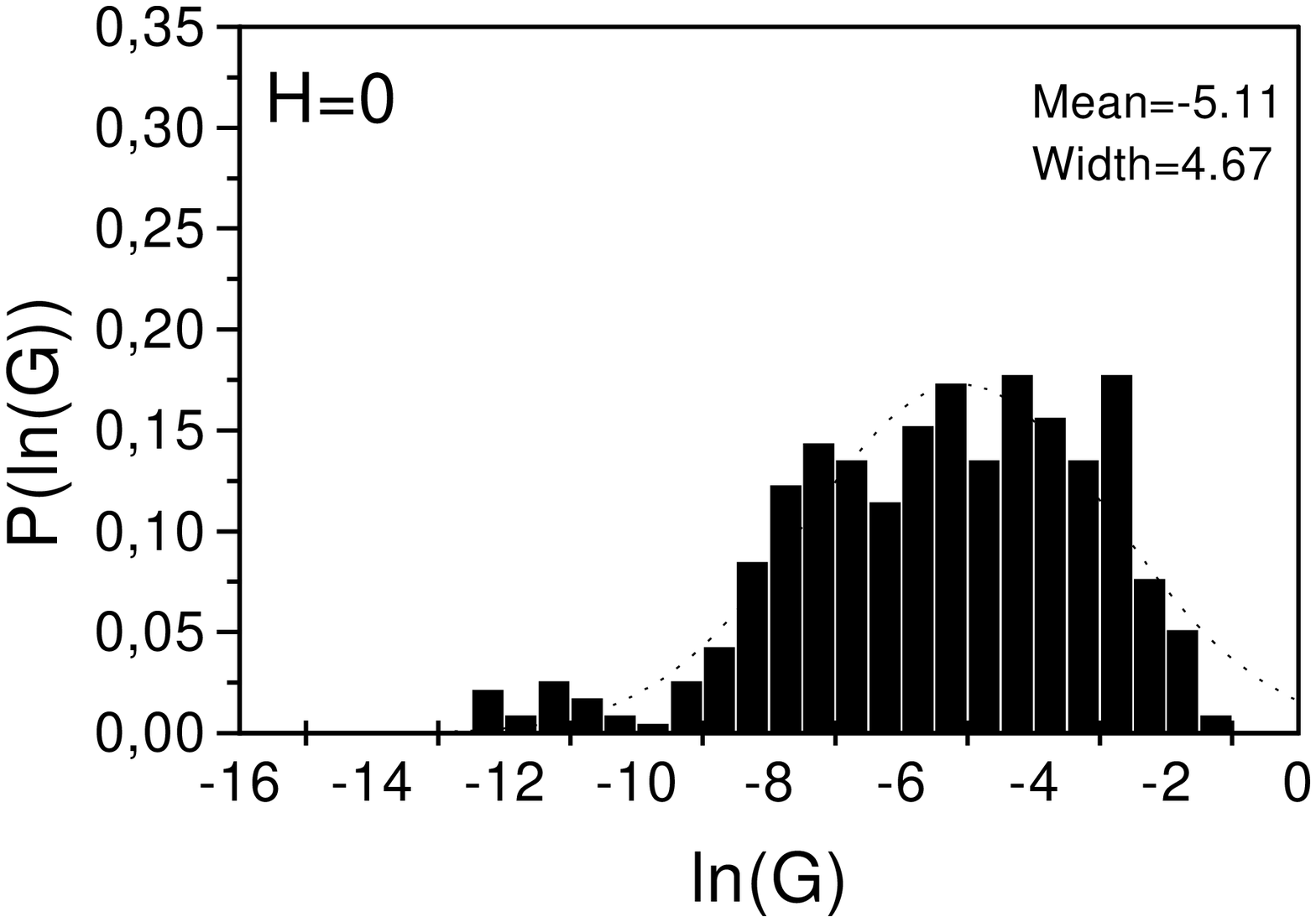,width=8cm}
 \epsfig{file=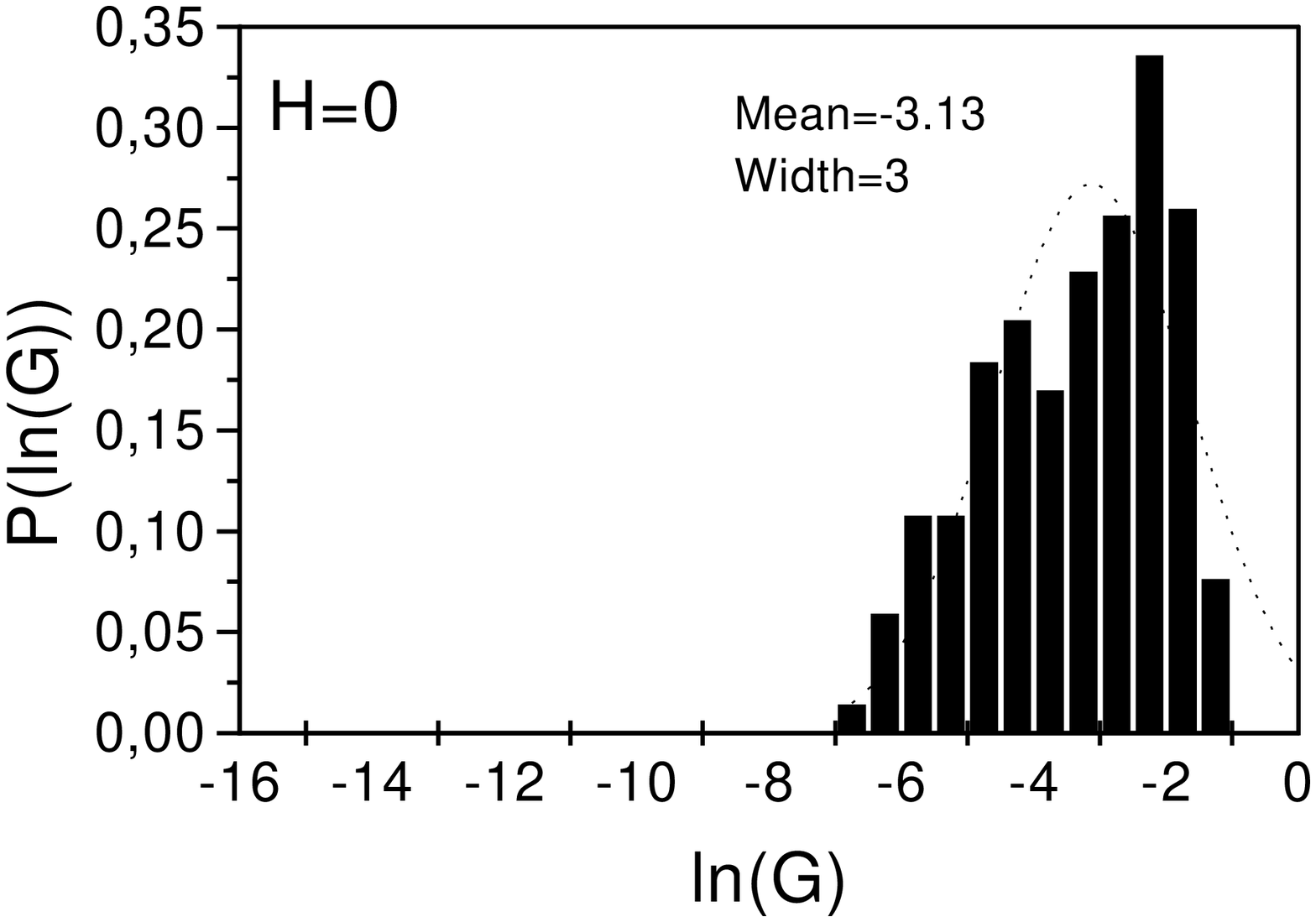,width=8cm}
\caption{A: Histogram of $\ln$G(T=100mK) for very negative gate voltages in the $0.5 \mu m$ long sample. B:Histogram of $\ln$G(T=100mK) for moderate negative gate voltages in the $0.5 \mu m$ long sample. C: Histogram of $\ln$G(T=100mK) for small negative gate voltages in the $0.5 \mu m$ long sample. Compare the shape of the distribution with the
figure 4 of reference \protect\cite{slevin}. The dotted lines are gaussain fits with parameters (mean value and width) indicated in the figures.}
\label{fig:histog}
\end{center}
\end{figure}

 The histogram of the conductance is shown for various gate voltage ranges on figure \ref{fig:histog}. In the deep localized regime the distribution shows a long tail to the high conductances, in contrast to the situation near the transition, where the distribution has a tail down to low conductances.

 \par These changes are not due to a  dimensionality crossover of the sample with gate voltage. A long tail to high conductances in the distribution is attempted  in the different context of the transerve conductance of a thin film or barrier \cite{hughes}. In that case the conductance is dominated by "punctures". This is the conjugated case of incoherent mesocopic effects in the 1D VRH case, where dominating resistances (~in series~) are now replaced by dominating conductances (~in parallel~). Wide and short geometries have been often used to study VRH regime \cite{popovic}\cite{xu}.
\par  The situation is completely different in our case where the sample is small in all dimensions. Eventually the sample is  more 1D at large negative gate voltages. The shape of the distribution deep in the insulating regime is due to the occurence of resonances. On the contrary, we believe that the shape of the distribution near the transition has better to be compared with zero temperature simulations, which show indeed a tail to low conductances near the MIT (~see for instance the figure 4 in reference \cite{slevin}~).This is not surprizing because our short sample is close to the zero temperature limit: there are resonant conductance peaks, the MIT occurs at $G \simeq {e^2\over h}$, the amplitude of the fluctuation corresponds to its predicted zero temperature limit (~see next section~).

\par Our pattern of conductance versus gate voltage is also reminiscent of the low temperature simulation of  1D disordered system by Azbel et al. \cite{azbel}. Very large peaks of conductance are superimposed on a background of a low  monotonic variation of $\ln (G)$ versus gate voltage, which is due to the decrease of $\left\langle \xi \right\rangle$ with $V_{G}$. 
\par As noted before, the conductance increases as temperature decreases for few conductance peaks. But other peaks show the opposite or even a non-monotonic behavior in temperature. This point has been already noted in ref. \cite{meir}
and attributed to a disorder effect. With some systematicity in our sample, we note that the higher peaks show $\delta G / \delta T \le 0$ in all the temperature range between the lowest temperature and 7 kelvin (~see figure   \ref{fig:gdet}~), intermediate peaks show $\delta G / \delta T \ge 0$ below $T \simeq T_{c}$ and $\delta G / \delta T \le 0$ above, and finally the lowest peaks show $\delta G / \delta T \ge 0$ in all the temperature range. We suggest that the transport through the whole sample could be partly resonant, partly hopping, and the balance between the two temperature dependences could explain the observed behavior.

\section{quantitative  estimations for the distribution of $\ln(G)$}

Let us become more quantitative on the amplitude of the conductance fluctuations.
Apart the simulations in 1D disordered wires, it is possible to get numerical results in higher dimension and even to get prediction from random matrices theory (~RMT~).  3D networks of 1D wires have been numerically studied by Avishai et al. \cite{avishai}. At zero temperature, the pattern of conductance versus energy consists in few resonant peaks superimposed on a smooth exponentially small background (~in the deep localized regime~). Also the second moment of the lognormal distribution of conductances obeys to:
\be var(\ln (G)) = -2\left\langle \ln (G) \right\rangle
\ee
the result attempted in quasi-one dimensional systems from RMT (~the factor of 2 disappears in 2 and 3d~).
\par
\begin{figure}[ptb]
\begin{center}
\epsfig{file=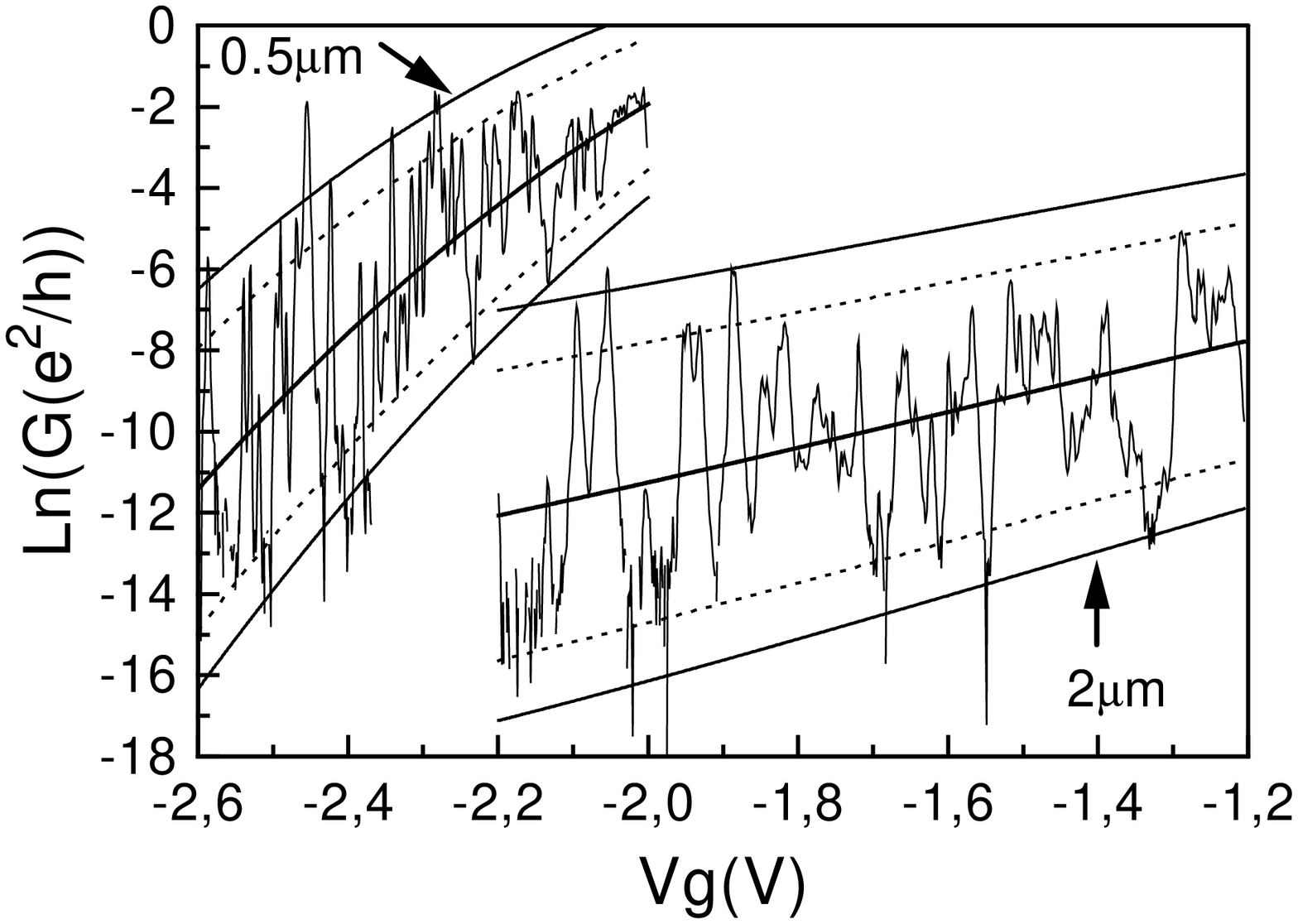,width=9.5cm}
\caption{$\ln$G(Vgate) at T=100mK for samples of lengths $0.5 \mu m$ and$2.0 \mu m$. Lines represent  the zero temperature estimation for the mean amplitude for the fluctuation of $\ln$(G), equation 9 (~see the text~) with a factor 2 (~solid lines~) and  without factor 2 (~dashed lines~) . This estimation  is well obeyed for the shortest sample (~$L = 0.5 \mu m$~) and the observed amplitude is slightly reduced for the $2 \mu m$ long sample.}
\label{fig:truncation}
\end{center}
\end{figure}

\begin{figure}[ptb]
\begin{center}
 \epsfig{file=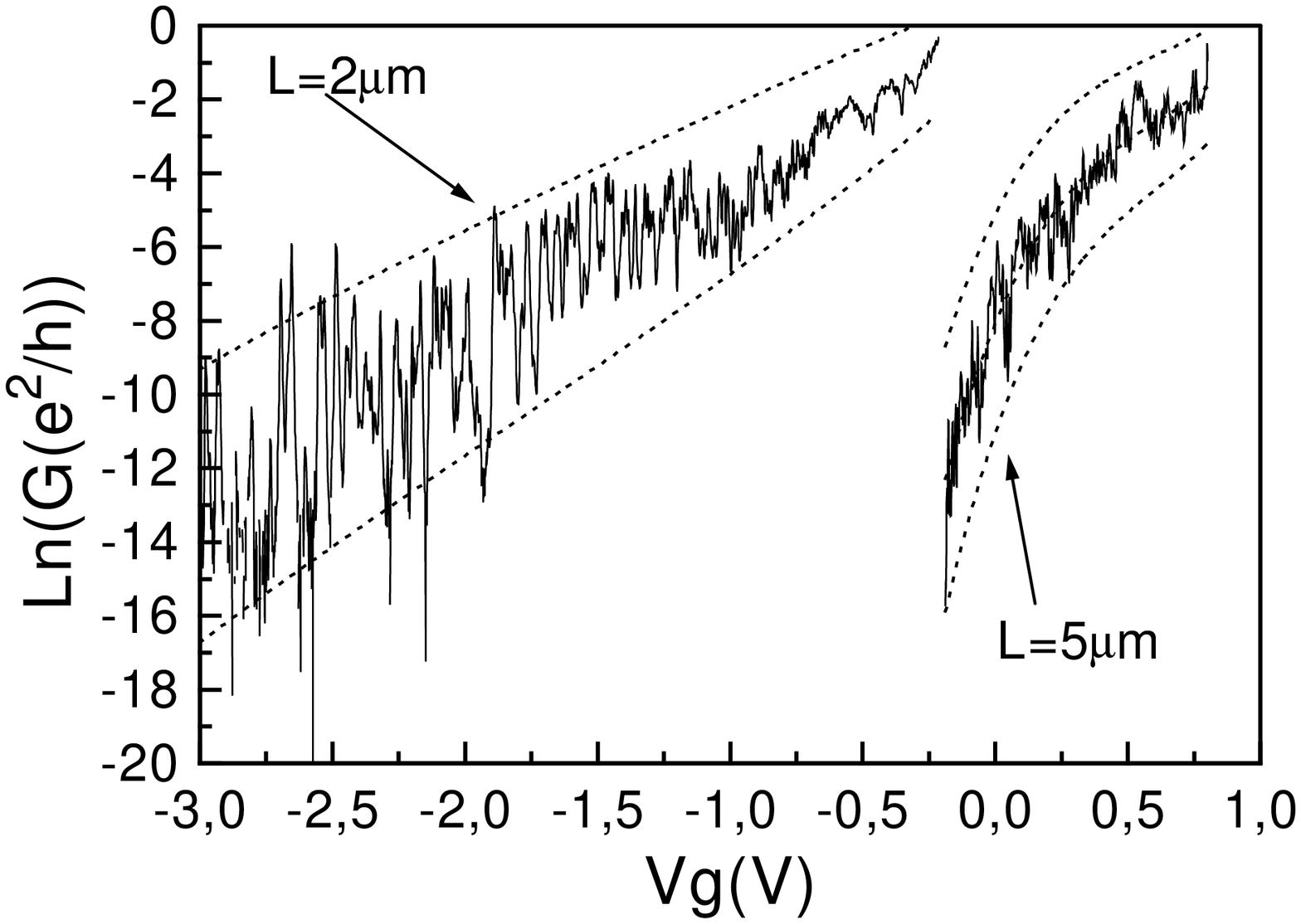,width=9.5cm}
\caption{$\ln$G(Vgate) at T=100mK for samples of lengths $2.0 \mu m$ and $5.0 \mu m$. Dotted lines represent  the zero temperature estimation for the mean amplitude for the fluctuation of $\ln$(G) (d=2,3). For the longer sample the quantum fluctuation is strongly truncated by geometrical fluctations \protect\cite{ladieu}. The gate voltage range is arbitrarily shifted from sample to sample.}
\label{fig:truncationL}
\end{center}
\end{figure}

 The amplitude of the conductance fluctuations in our short sample is indeed very close to this estimation. Figures \ref{fig:truncation} and  \ref{fig:truncationL} show what is typically attempted with the help of equation (9).
The solid lines represent the estimation $ \left\langle \ln(G) \right\rangle +- \sqrt{ -2 \left\langle \ln (G) \right\rangle}$ and the dashed lines: $ \left\langle \ln (G) \right\rangle +- \sqrt{ - \left\langle \ln (G) \right\rangle}$. Contrarily to the good accordance with equation (10) obtained in the short sample (~$L = 0.5 \mu m$~), it is clear that the fluctuation is reduced in longer samples (~$L = 2.0 \mu m$ and $L = 5 \mu m$~ ). This effect is due to the truncation by the geometrical fluctuations as described by Ladieu et al. \cite{ladieu}. It has been shown that for wires whose length is $N$ times the
mean Mott hopping length, the amplitude of the fluctuation is given by 
$ { \Delta \ln (G) \over \ln (G) } \simeq { 1\over 2 \ln (aN)}$, where $a$ is a numerical constant of order of 2 \cite{ladieu}. This is obeyed for our long samples \cite{ladieu}.

\par Note that he RMT result (~eq. (9)~) is obtained after averaging over disorder. Varying the energy with the gate is not equivalent to average over disorder, because the continuous variation of the gate voltage matches any resonance (~of finite width at non zero temperature~), even if their density is extremely small.
\par Mucciolo et al. have performed simulations on a 2D Anderson hamiltonian \cite{jalabert}. The simulation is done on a rectangular lattice of $34 \times 136$ sites with various on-site disorder. They obtain  the conductance pattern versus energy. In the strongly insulating regime (~large disorder~), the transport is dominated by resonant tunneling. A model based on a simplified description of the localized wave functions (~uniformly distributed in the sample, with inverse localization lengths following a Gaussian distribution~) explains their numerical results. The value of the conductance at the resonance depends on the coupling to reservoirs which is largely fluctuating (~see also ref. \cite{azbel}~).
\par We will use the correspondance between conductance resonances and energy states obtained in simulations \cite{jalabert} to address the question of energy level statistics.

\section{Spacing rigidity}

Our samples have the interesting property that they are {\it small in all dimensions}, in contrast to many studies where the cross-section is large. Near the transition, the cross-section is less than the localization length. This means that if we neglect interaction effects, as we have done up to now, the position of the conductance resonances in energy should exhibit some spacing rigidity. This is a consequence of the spacial overlapping between Anderson localized wavefunctions which are rather extended.
\par As noted before, no spacing rigidity is attempted if the wavefunctions are strongly localized or if the sample volume is much larger than the Mott hopping  length. On the other hand, if Coulomb interaction enters into the game, the situation changes:
periodicity in gate voltage can appear in the Coulomb blockade regime for instance \cite{staring}.
\par
\begin{figure}[ptb]
\begin{center}
 \epsfig{file=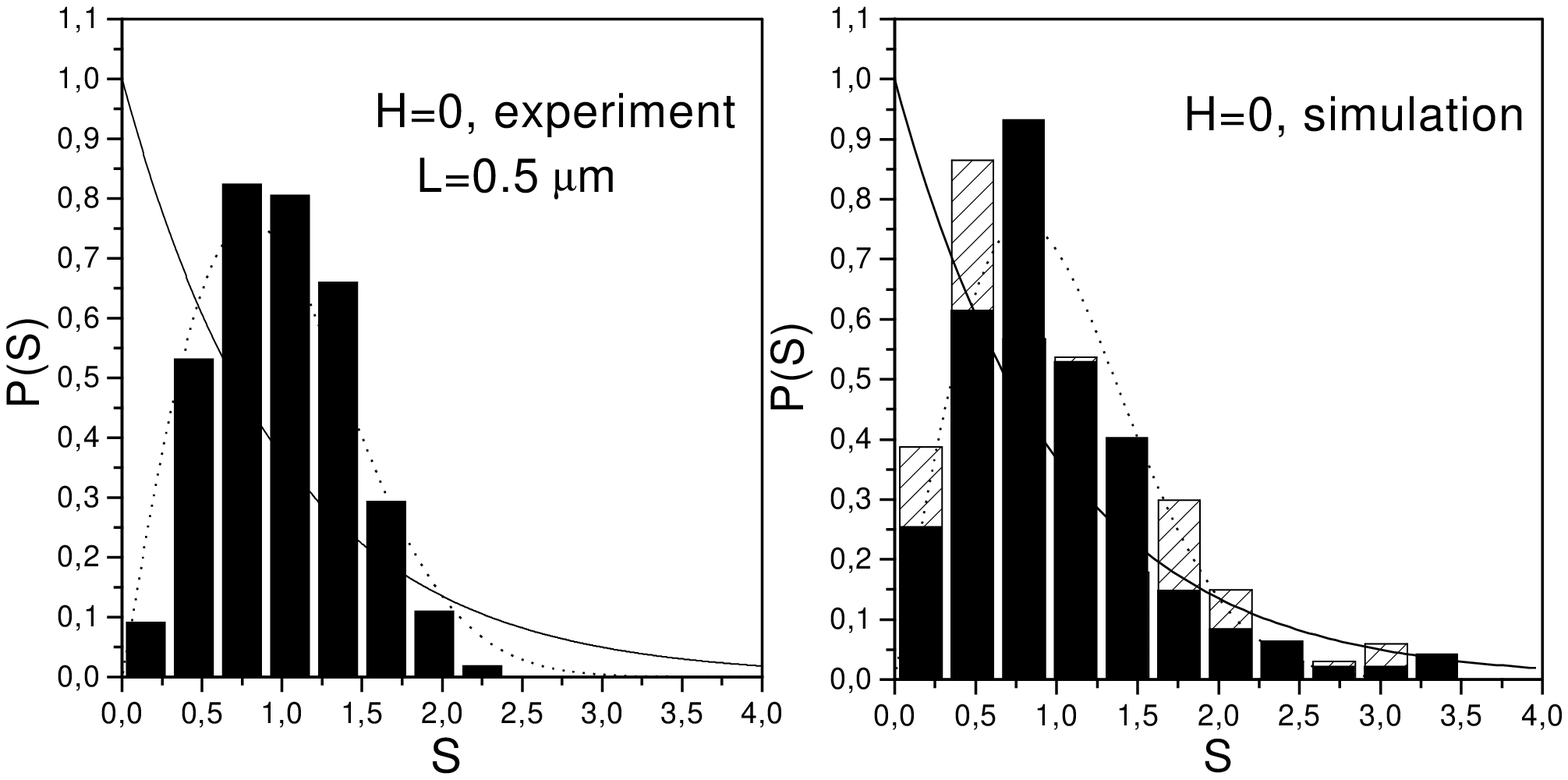,width=14cm}
\caption{Left pannel: Spacing distribution for conductance resonances. Each resonance is considered whatever the value of the conductance maxima. The spacing is normalized to the mean spacing (~7 mV~). The statistics is over 180 resonances on the same sample for 3 annealings, which decorrelate completely the conductance pattern. The Poisson behavior (~solid line~) and the Wigner-surmise (~dotted line, $\beta = 1$~) are shown for comparison (~see the text~). By principle p(s=0) = 0, such that the measured distribution should deviate from Poisson in any case. However the accordance with WS, particularly at large s is excellent. Right panel: The same distribution, calculated from the conductance data simulated by Mucciolo et al. (~same mathematical treatment~) is shown for comparison. The black and dashed bars correspond to two values for the disorder parameter. The simulation is done on a rectangular lattice of $34 \times 136$ sites with various on-site disorder (Anderson Hamiltonian).\protect\cite{jalabert}.}
\label{fig:histo}
\end{center}
\end{figure}

 Figure \ref{fig:histo} shows the nearest spacing distribution over more than 180 conductance resonances in our shortest sample for 3 annealings to room temperature. It integrates all the resonances in the insulating regime for various gate voltages. Each maximum of $\ln (G)$ is considered as a resonance whatever is the value of $\ln(G)$ at the resonance. By  principle of the determination: p(s=0) = 0. The mean spacing between resonance is approximately constant (~7 mV~) in gate voltage and agrees well with our estimation for the mean 
level spacing (~see section 1~).
\par Our measured spacing distribution is very close to the Wigner-surmise distribution  (~$ p(s) = ({\pi \over 2}) s^{\beta} \exp (- \pi s^{2}/ 4)$ (~$\beta= 1$~) and $ p(s) = ({32 \over \pi^{2}}) s^{\beta} \exp (- 4 s^{2}/ \pi)$ (~$\beta= 2$~)~) and strongly departs from the Poisson behavior (~$ p(s) = \exp (-s)$~). Small spacings are not really discriminant because by the principle : p(s=0) = 0. Large spacings are perhaps more important to discriminate both behaviors: our observation indicates a large deviation from Poisson law.
\par In the disordered metallic regime, The Wigner surmize is attempted  for the level distribution. It has been observed both by capacitance measurements \cite{ashoori} or by conductance measurements using  resonant tunneling  through one electrode \cite{sivan}. Some rigidity should persist in the insulating regime close to the MIT. The level and conductance statistics precisely at the MIT has been the subject of intense theoretical  work (~see for instance the reference \cite{kramer} and references inside~). In the insulating case, the localized states are weakly coupled to the reservoirs (~the electrodes~), such that the conductance resonances distribution reflects directly the level spacing distribution (~neglecting interactions~).
Again we observe a very good similarity between our experiment and independent electron models.
\par
\begin{figure}[ptb]
\begin{center}
 \epsfig{file=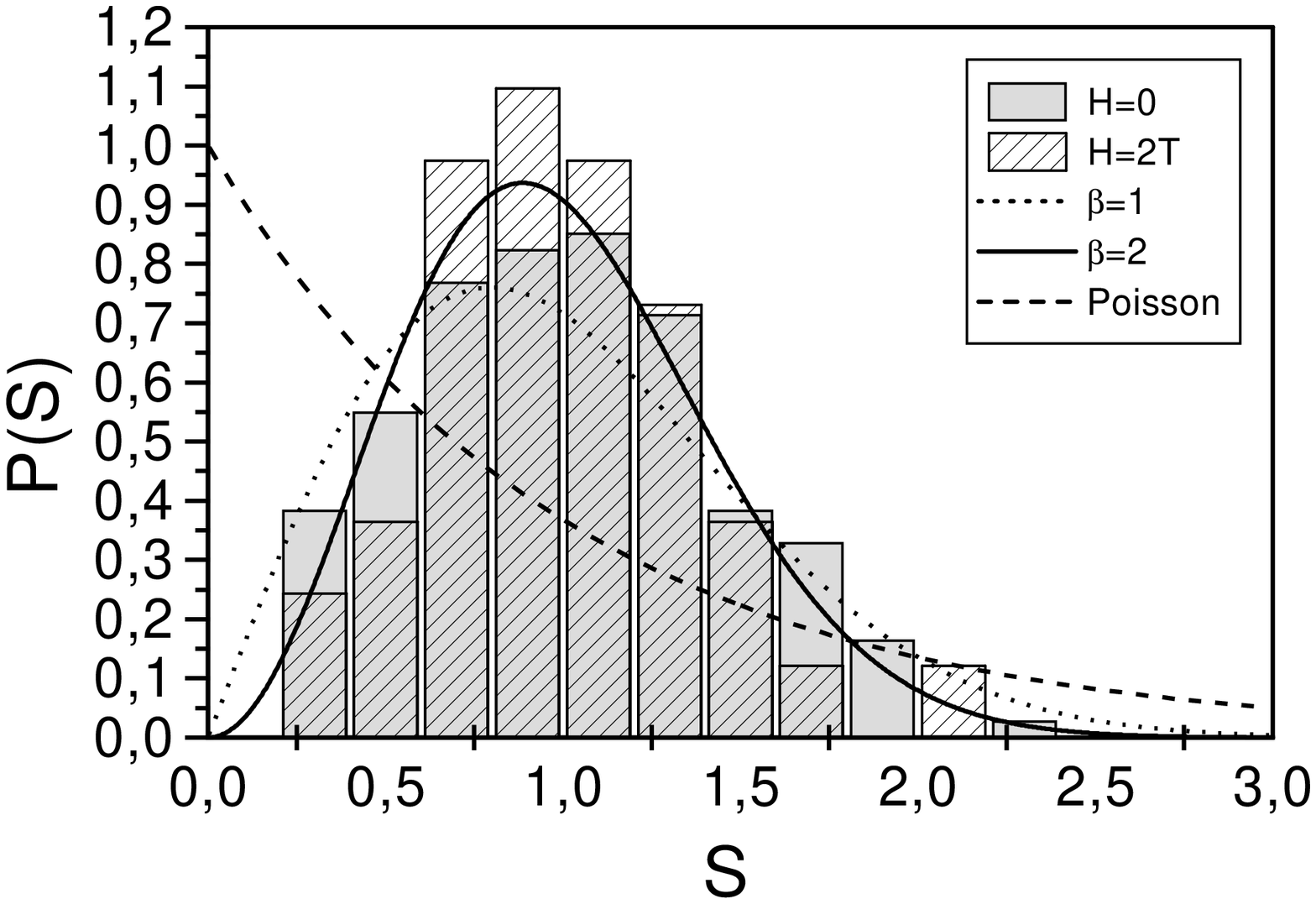,width=14cm}
\caption{Spacing distribution for conductance resonances at H=0 and H=2T.
 The rigidity is slightly reinforced in magnetic field what is attempted for the  $\beta = 1$ to  $\beta = 2$ transition for the Wigner surmise.  The mean spacing is not changed with H. The statistics is over 50 resonances only at H=2T. The Poisson behavior (~dashed line~)  the Wigner-surmise (~dotted line: $\beta = 1$, solid line:  $\beta = 2$~)  are shown for comparison.}
\label{fig:histoh}
\end{center}
\end{figure}

 To reinforce our observation we have made the same statistic in a magnetic field of 2 teslas. Unfortunately only 50 resonances have been recorded.
Figure \ref{fig:histoh} shows that the rigidity is somewhat reinforced as compared to the zero magnetic field case. This reinforcement is also observed in the simulation of Mucciolo et al. \cite{jalabert}. The statistic is not sufficient to insure that the $\beta= 1$/$\beta= 2$ transition is seen by time reversal symmetry breaking for the Wigner surmize distribution. Surely such an observation would be a crucial test for the proposed interpretation.
\par
\begin{figure}[ptb]
\begin{center}
 \epsfig{file=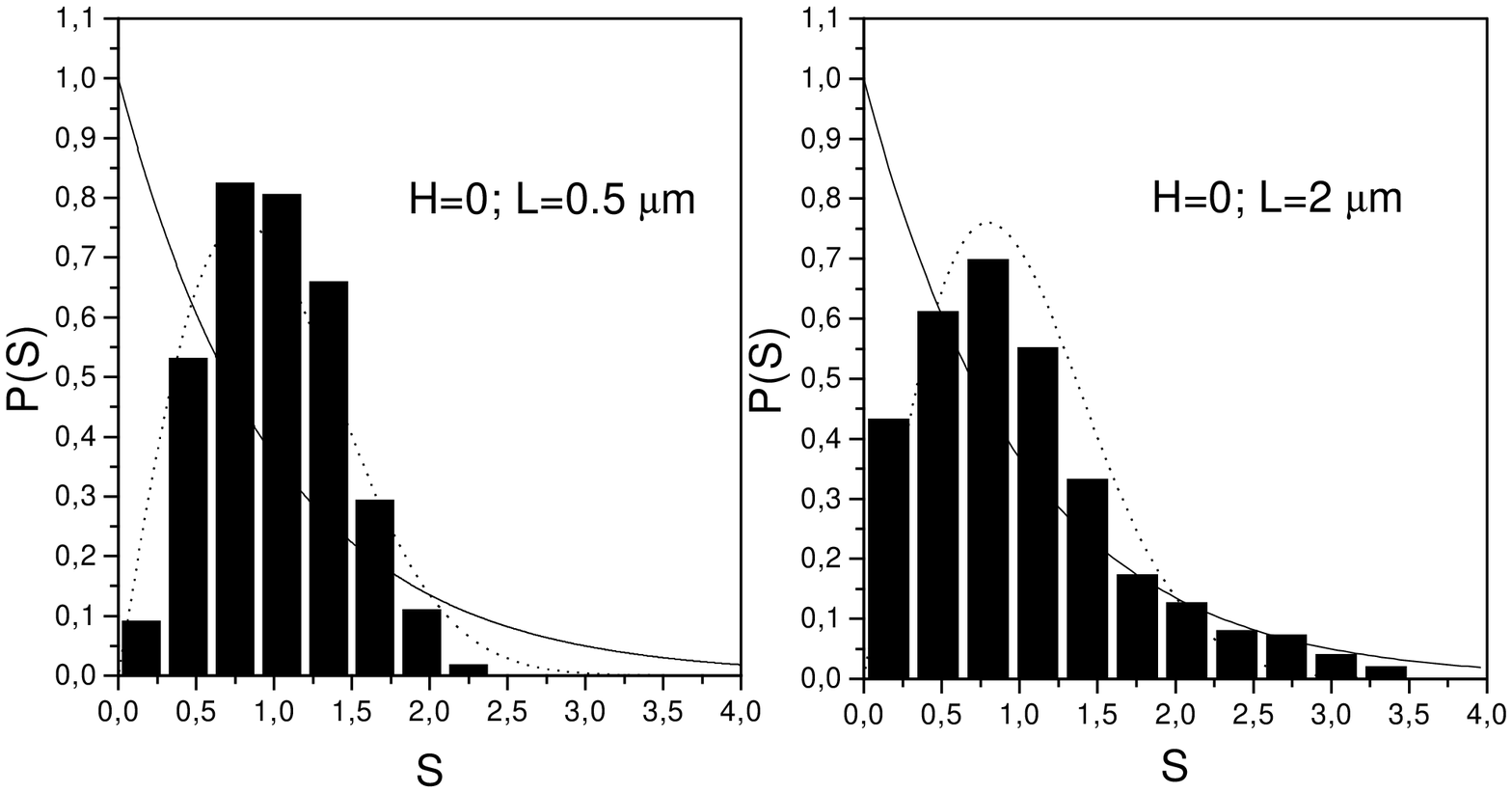,width=10cm}
\caption{Spacing distribution for conductance resonances for the $L=0.5 \mu m$  (left pannel) and $L=2 \mu m$ samples (right pannel).  Again for the $L=2 \mu m$ sample a large deviation from Poisson  is found, but the accordance with Wigner Surmize is less good than in the shortest coherent sample
 particularly for large spacings, where it follows a poisonnian behavior.}
\label{fig:histoL}
\end{center}
\end{figure}

 As said before the spectral rigidity should weaken if  the sample size increases. For instance, if its length increases to be much larger than the Mott hopping length, the spacing distribution should reflects the Poissonian behavior attempted for incoherent geometrical mesoscopic effects.
Figure \ref{fig:histoL} shows indeed that for a $ 2 \mu m$ long sample (~500 peaks recorded~) the fit to the Wigner surmize is less good than for the shortest sample. As noted before, p(s=0)=0 by principle such that
the measured distribution should deviate from Poisson in any case. But at large spacings, p(s) follows Poisonnian behavior for the long sample and Wigner -surmize for the short sample.

\section{magnetoconductance}

Magnetoconductance in disordered insulators is a very debated point. Important results have been already obtained in long quasi 1D samples in the variable range hopping regime.  Large fluctuations in gate voltage are due to changes of the dominating resistor in a sequence of inelastic hops \cite{lee}. Large fluctuations of the conductance with
magnetic flux are due to the interference effect described first by Nguyen et al \cite{nguyen}.
\par
\begin{figure}[ptb]
\begin{center}
\epsfig{file=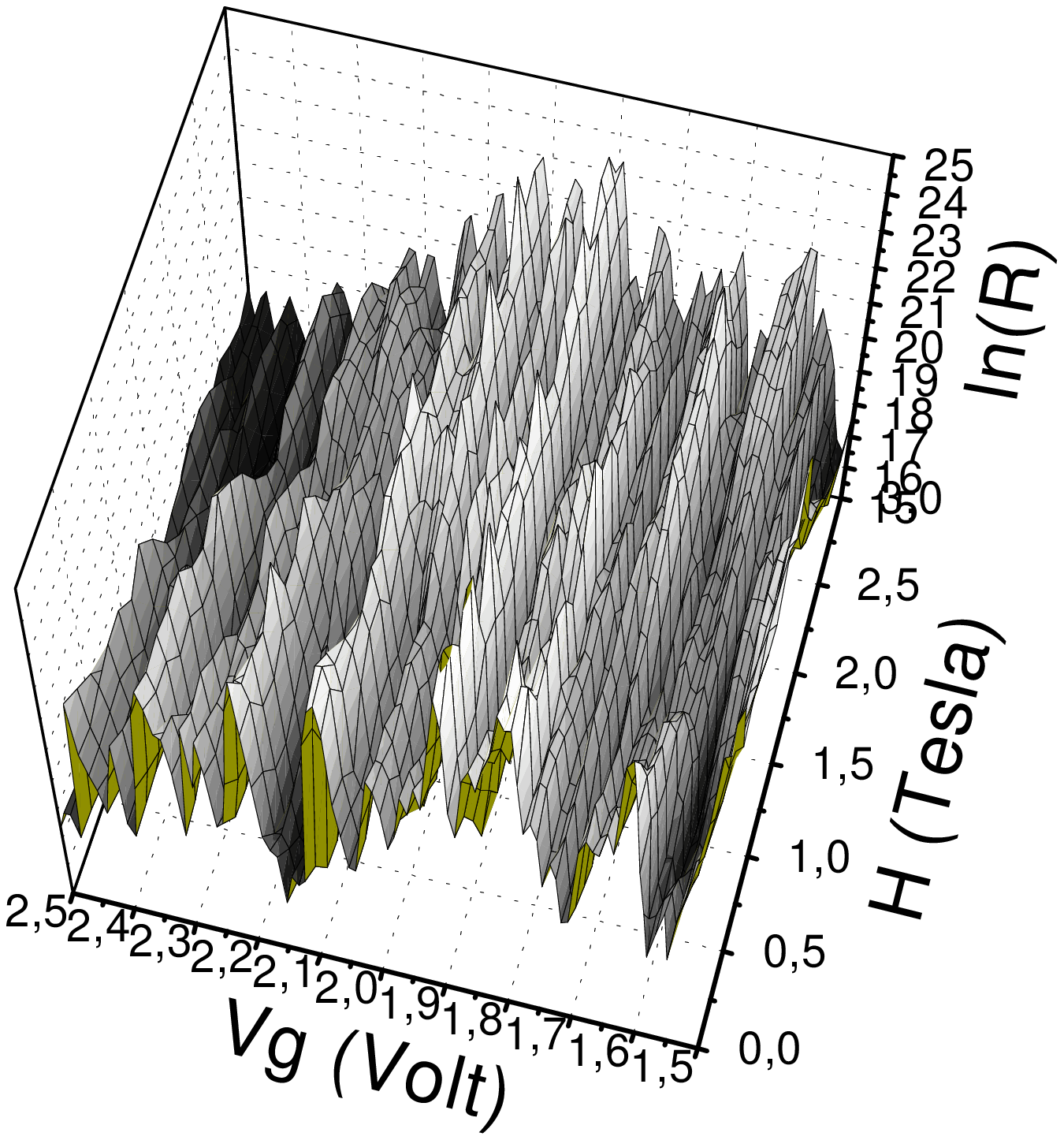,width=10cm}
\caption{$\ln$R(Ohms) versus H and versus $V_{G}$ at T=100 mK in the in the $2 \mu m$ long sample
in the strongly localized regime (absolute values for negative gate voltages are plotted). Note the basic non-ergodicity in magnetic field.}
\label{fig:vgeth}
\end{center}
\end{figure}

 The magnetic field changes the interference pattern due to intermediate impurity diffusion on the dominating hop. The Nguyen-Spivak-Shklovskii (~NSS~) model explains the large magnetoconductance fluctuations seen deep in the insulating regime \cite{raikh}\cite{nguyen}, and also why the interference effect 
does not decorrelate however the pattern of fluctuation in energy (~non ergodicity~). Non ergodicity  is not a proof in favor of Coulomb blockade interpretation  in insulating wires. The non-ergodicity is
illustrated on figure \ref{fig:vgeth}, obtained for  the $ 2 \mu m$ long sample
in the deep insulating regime. This
proves that Zeeman effects, which imply ergodicity, are not important in GaAs contrarily to the case of
Silicon \cite{ladieu}.
\par In the following we will show that independent electron models explain the mean magnetoconductance behavior observed in our experiment both in the strongly localized regime and near the transition. The mean magnetoconductance effect is not considered in models based on the Coulomb blockade interpretation.
\par In the NSS model the mean magnetoconductance is given by:
\be \left\langle \ln G(H=2T) \right\rangle - \left\langle \ln G(H=0) \right\rangle \simeq 1
\ee  (~in quantum units~), at least for Mott hopping length not too large as compared to the localization length. This gives a relatively small positive magnetoconductance, rather independent of the value of $\left\langle \ln G(H=0) \right\rangle$ itself. More recently, it has been suggested that NSS model gives an effective increase of the localization length, which produces a much larger positive  (~and $\left\langle \ln G(H=0) \right\rangle$ dependent~) magnetoconductance than the original estimation (equation 10) \cite{spivak}. The NSS model is based on the assumption that interference effects due to intermediate impurity scattering during tunneling are important only between forward-directed pathes. Essentially one neglects pathes which include loops, that is valid if the localization length is smaller than the distance between impurities. But it is impossible to neglect the loops when the localization length becomes larger than $\ell$.  A doubling of the localization length is predicted in quasi 1D geometry for a large number of channels (~without spin-orbit scattering~)\cite{pichard}. The effect is due to time reversal symmetry breaking on the scale of $\xi$. Bouchaud et al. \cite{bouchaud} have shown how interferences amongst the diffusive loops inside the localization domain produce the effect.
 Near the MIT in large GaAs insulating wires, a doubling of the
localization length has been observed recently by applying a magnetic field \cite{gershenson}. This confirms results previously obtained  in other disordered insulators \cite{pichard}.  Also near the transition the fluctuations in magnetic field become ergodic \cite{ladieu}.
\par A very important point is to understand how to conciliate the NSS model in the strongly localized regime (~$\xi \le \ell$~), which predicts no or small increase of $\xi $, and the prediction in the barely localized regime (~$\xi \ge \ell$~), i.e. the doubling of $\xi$.  For the first time,  in our quasi 1D samples we are able to investigate both limits, by varying the gate voltage, i.e. the number of channels. 
\par We have already shown (~equations (5)(6)~) that there is only one canal left for $V_{g} = -2.4 V$. For this value, the localization length is reduced to twice the elastic mean free path $\ell$,
and there is no more symmetry dependence of $\xi$ (~equation (2) with ${\cal N}=1$~). This case corresponds to the strongly localized regime where the NSS assumption is more valid. For
 $V_{g} = -1.0 V$ on the contrary, ${\cal N}=$=8-10 and $\xi$ is much larger than $\ell$:
$\xi \simeq (N + 1)\ell \simeq 0.18 \mu m$. We attempt an approximate doubling of $\xi$ .

\par With the help of equations (1) and (2) one gets:
\be
{\left\langle \ln G(H=0) \right\rangle -  \left\langle \ln G(H=2T) \right\rangle \over \left\langle \ln G(H=0) \right\rangle} = 1 - {\xi(\beta = 1) \over \xi(\beta = 2)} = {1 \over 2} ( 1 - {1 \over {\cal N}}) = {1 \over 2} ( 1 + { \ell \over L} \left\langle \ln G(H=2T) \right\rangle) 
\ee
i.e. the mean relative magnetoconductance is {\it proportionnal} to $\left\langle \ln G(H=2T) \right\rangle$ (~the exact determination for the number of channels is eliminated in this procedure~). The
prefactor outside the exponential have not been considered here.
The value of $H=2T$ has been choosen because the low field positive magnetoconductance (~associated to the time reversal symmetry breaking $\beta =1 /\beta =2$ on
the localization length scale~) is already saturated at this value, and in addition the negative exponential magnetoconductance due to the shrinking of
Bohr orbitals appears only above $H=4T$. Between $H=2T$ and $H=4T$ there is a plateau in the magnetoconductance (~not shown~).

\par
\begin{figure}[ptb]
\begin{center}
\epsfig{file=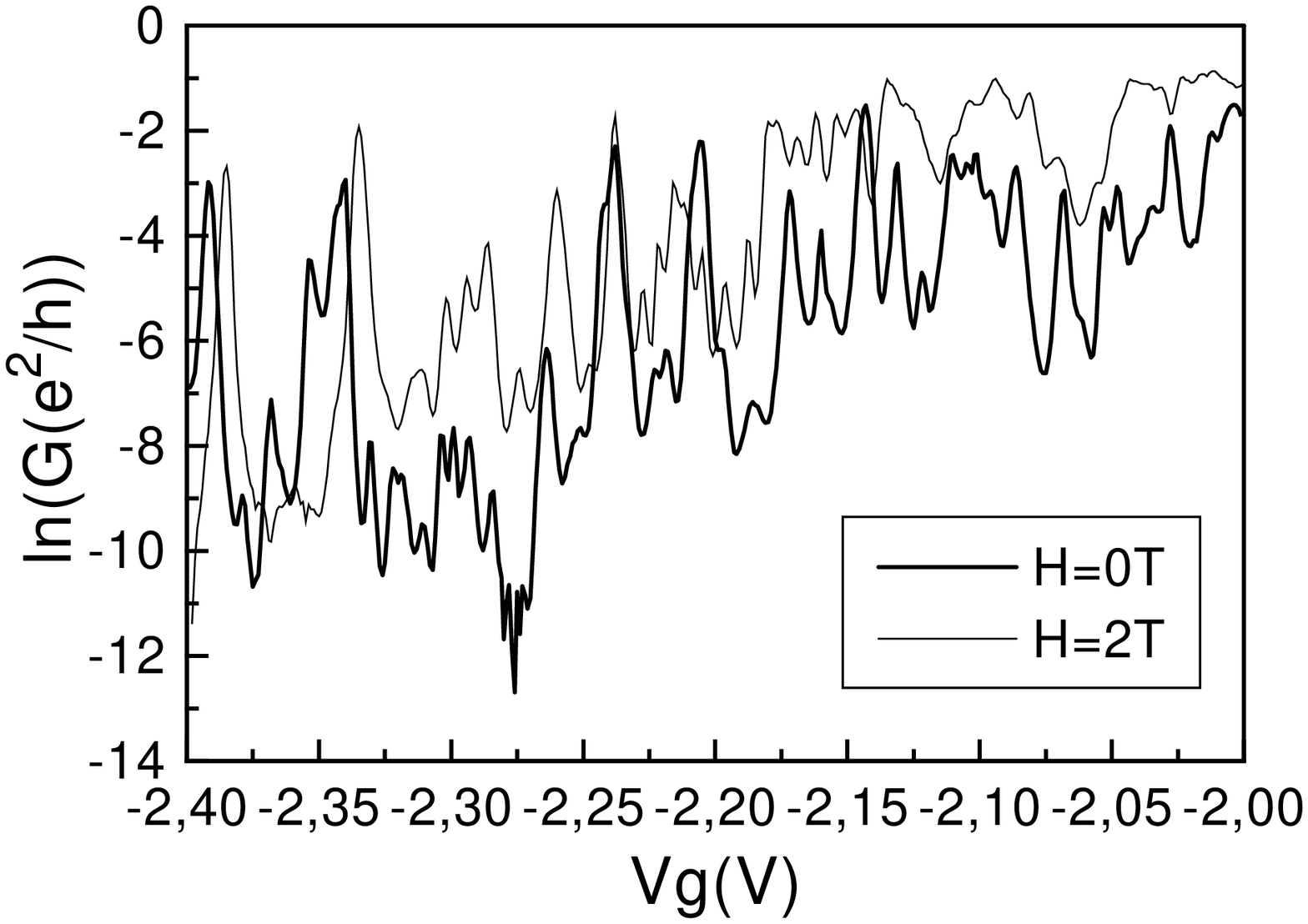,width=8cm}
\epsfig{file=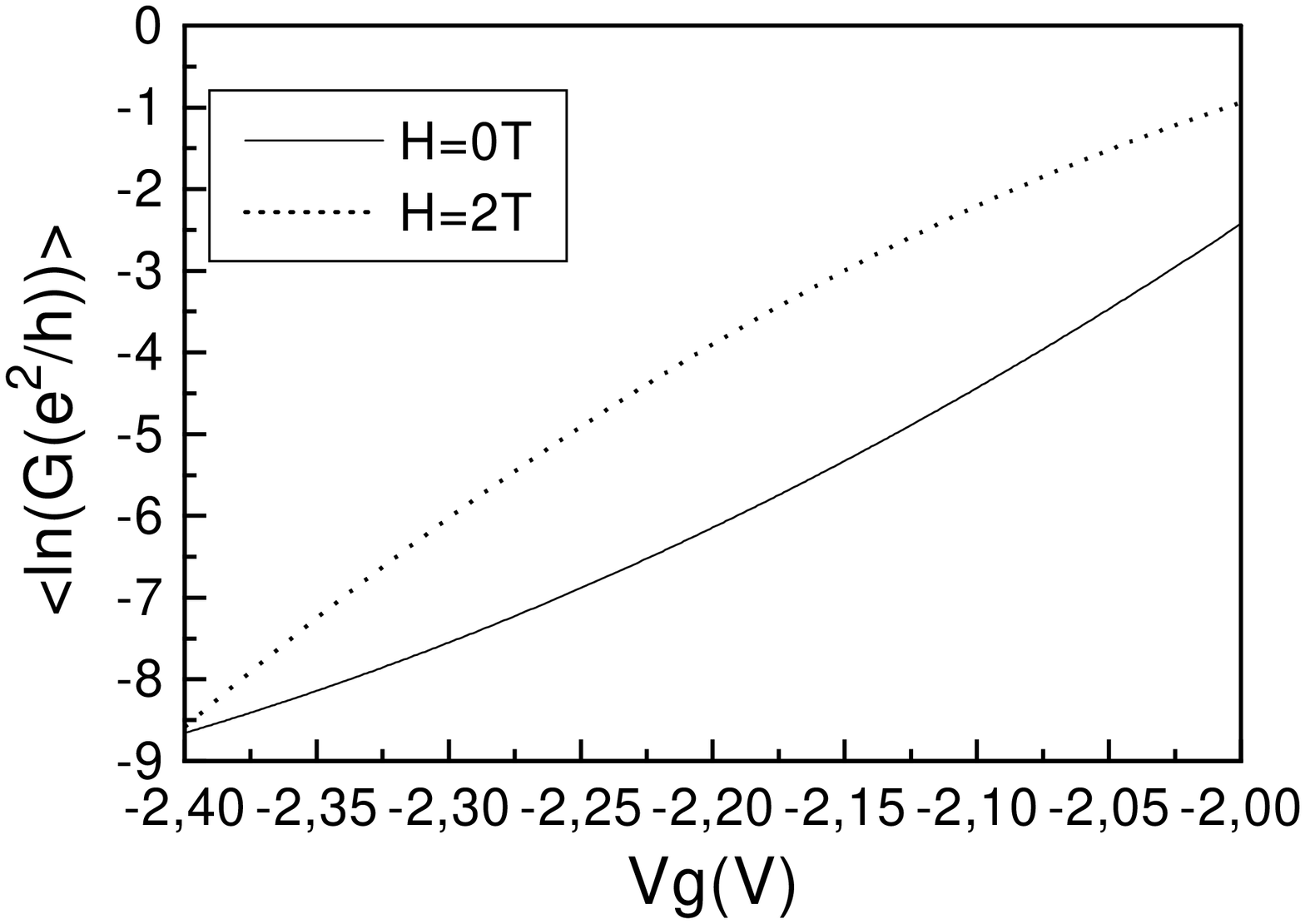,width=8cm}
\epsfig{file=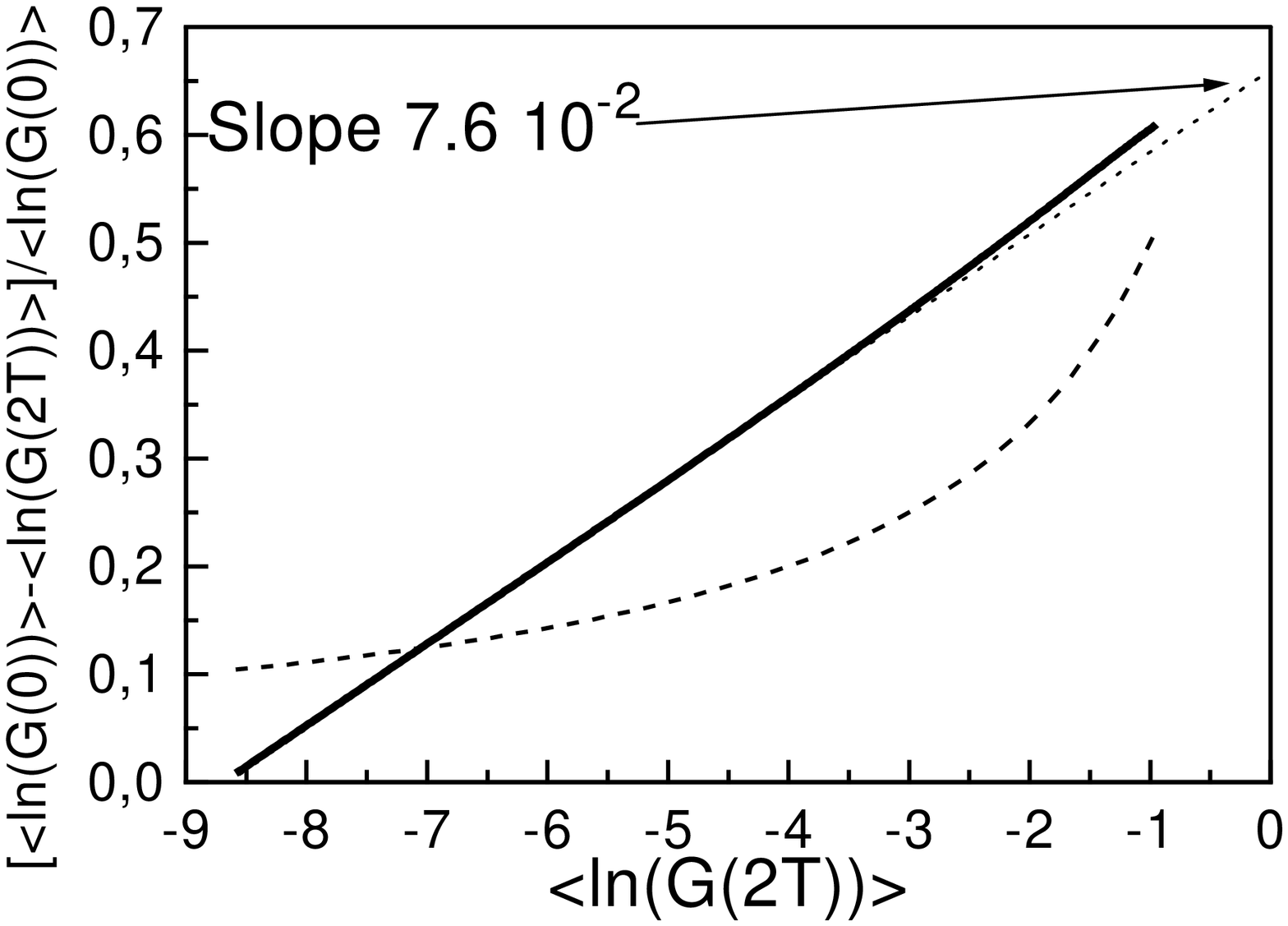,width=8cm}
\caption{ A: $\ln$(G) versus gate voltage for two magnetic fields at $T \simeq 100 mk$ in the $L=0.5 \mu m$ sample. B: The numerically averaged ln(G) versus gate voltage. C: The mean positive magnetoconductance $(\left\langle \ln G(H=0T) \right\rangle-\left\langle \ln G(H=2T) \right\rangle)/\left\langle \ln G(H=0T) \right\rangle$
versus $\left\langle \ln G(H=2T) \right\rangle$ at T=100mK. The dotted line is the fit by equation (11), 
i.e. the Random Matrix Theory estimation by Pichard et al. \protect\cite{pichard}. The dashed line is the estimation by equation (12)}
\label{fig:MC}
\end{center}
\end{figure}

  Figure \ref{fig:MC} shows indeed that equation 11 is perfectly obeyed. The slope is $ 7.6 \ \ 10^{-2} = { \ell \over 2L}$ that gives $L \simeq 160 nm$ (~$\ell \simeq 20 nm$~). $L$ is comparable to our estimation of the phase breaking length. Deviation at high conductances could be attributed to the neglected prefactor dependence of $\ln(G)$ on $N$.
\par On figure \ref{fig:MC} we have also plotted the results of equation (11), i.e. the original estimation by NSS:
\be {\left\langle \ln G(H=0) \right\rangle - \left\langle \ln G(H=2T) \right\rangle \over \left\langle \ln G(H=0)\right\rangle} = {1 \over 1 - \left\langle \ln G(H=2T) \right\rangle} 
\ee
In the strongly localized regime (large negative $\left\langle \ln G(H=2T)\right\rangle$ values), both estimations give similar results, compatible with the observation, but for intermediate values of $\left\langle \ln G(H=2T) \right\rangle$, we observe a strong deviation from the NSS estimation.
\section{conclusions}
 We have studied very small GaAs MESFETs at very low temperature, and obtained an elastic transport regime for a disordered insulator of size $70 nm \times 70nm \times 0.5 \mu m$. The analysis of the temperature and energy dependence of the conductance resonances proves that the transport is dominated by elastic tunneling. The nearest energy spacing statistics of conductance resonances shows some rigidity , which is attributed to the localized levels statistics itself. The histogram of conductances has been also studied. Near the metal-insulator transition, it exhibits a long tail to low conductances. The variance of the distribution is in accordance with the zero temperature random matrix theory (RMT) prediction in the insulating regime.
The mean magnetoconductance has been studied as a function of the number
of channels in the insulating regime. It is also in accordance with the RMT prediction for quasi 1D samples without spin-orbit scattering. When the number
of channel is large, one get a doubling of the localization length by application of a magnetic field, but as this number becomes comparable
to 1, the positive magnetoconductance becomes much less important.
\section{aknowledgments}
We thanks B. Etienne for providing the MBE GaAs:Si layers and R. Tourbot
 for its technical help. This work is partly supported by the INTAS contract 94.4435.



\end{document}